\documentclass[10pt,final,times,journal,twocolumn]{IEEEtran}
\usepackage[normalem]{ulem}
\usepackage[hyphens]{url}
\usepackage{microtype}

\usepackage[usenames,dvipsnames,table]{xcolor}
\usepackage{tcolorbox}
\usepackage{framed}
\usepackage{footnote}
\usepackage{graphicx}
\usepackage{subcaption}
\usepackage{pifont} 
\usepackage{tikz}

\usepackage{enumitem}

\usepackage{tabu}
\usepackage{booktabs}
\usepackage{multirow}
\usepackage{tablefootnote}
\usepackage[bookmarks=true,breaklinks=true,letterpaper=true,colorlinks,linkcolor=black,citecolor=blue,urlcolor=black,hyperfootnotes=false]{hyperref}

\title{Efficient Bypass in Mesh and Torus NoCs} 
\author{\IEEEauthorblockN{Iv\'an P\'erez\IEEEauthorrefmark{1},
					  Enrique Vallejo\IEEEauthorrefmark{1} and
                      Ram\'on Beivide\IEEEauthorrefmark{1}
        \thanks{This work was supported by the Spanish Ministry of Science, Innovation and Universities, FPI grant BES-2017-079971, the Spanish Ministry of Science, Innovation and Universities under contract TIN2016-76635-C2-2-R (AEI/FEDER, UE) and the European HiPEAC Network of Excellence. The Mont-Blanc project has received funding from the European Union’s Horizon 2020 research and innovation programme under grant agreement No 671697.}
		}\\
		\IEEEauthorblockA{\IEEEauthorrefmark{1}University of Cantabria\\
			Santander, Spain							
		}
}

\definecolor{colboxcolor}{HTML}{CEE3F6}

\definecolor{gray}{rgb}{0.5,0.5,0.5}
\definecolor{dollarbill}{rgb}{0.52, 0.73, 0.4}
\definecolor{bittersweet}{rgb}{1.0, 0.59, 0.54}
\definecolor{chromeyellow}{rgb}{1.0, 0.75, 0.3}

\newcommand{\true}{\textcolor{dollarbill}{\ding{51}}}
\newcommand{\false}{\textcolor{red}{X}}

\begin{document}

\maketitle

\begin{abstract}
Minimizing latency and power are key goals in the design of NoC routers. Different proposals combine lookahead routing and router bypass to skip the arbitration and buffering, reducing router delay.
However, the conditions to use them requires completely empty buffers in the intermediate routers. This restricts the amount of flits that use the bypass
pipeline especially at medium and high loads, increasing latency and power.

This paper presents \emph{NEBB}, Non-Empty Buffer Bypass, a mechanism that allows to bypass flits even if the buffers to bypass are not empty. The mechanism applies to wormhole and virtual-cut-through, each of them with different advantages. \emph{NEBB-Hybrid} is proposed to employ the best flow control in each situation. The mechanism is extended to torus topologies, using FBFC and shared buffers.

The proposals have been evaluated using Booksim, showing up to 75\% reduction of the buffered flits for single-flit packets, which translates into latency and dynamic power reductions of up to 30\% and 23\% respectively. For bimodal traffic, these improvements are 20 and 21\% respectively. Additionally, the bypass utilization is largely independent of the number of VCs when using shared buffers and very competitive with few private ones, allowing to simplify the allocation mechanisms.

\end{abstract}


\begin{tikzpicture}[remember picture,overlay]

\node[anchor=north,yshift=-10pt] at (current page.north) {

    \fbox{\parbox{\dimexpr.95\textwidth-\fboxsep-\fboxrule\relax}{
        This is a accepted manuscript of the paper; the final version of this work has been published in the Journal of System Architecture, DOI: https://doi.org/10.1016/j.sysarc.2020.101832.}

        }

    };

\end{tikzpicture}

\nointerlineskip

\section{Introduction} \label{sect:introduction}
NoC latency has a clear impact on memory access time, and thus on the system
performance.
One of its most critical components is the router delay, because of the
traditional pipelined implementation with several stages.

To minimize such latency, different mechanisms have been proposed to reduce the router
pipeline stages, including lookahead routing~\cite{Galles1997} and router bypass~\cite{kumar2007}.
Together, these mechanisms allow for a single-cycle router implementation, plus
one cycle for link traversal. Additionally, bypass mechanisms reduce the use of
buffers, which are the most power hungry components in the router. For these
reasons, maximizing the utilization of the bypass pipeline is important for NoC
latency and power reduction.

The bypass path is used when certain conditions (the \emph{bypass conditions)} hold,
as detailed in Section~\ref{sect:bypass_intro}. The bypass conditions used
in previous proposals~\cite{kumar2007}\cite{kumar2008} guarantee that packets do
not interleave in the same buffer (corrupting data) and, by construction, also preserve
order for packets sent in the same path and virtual channel (VC), even when they do not belong
to the same flow. Message ordering is not a requirement of many coherence protocols
(or, at least, not in all the virtual networks) and it is in fact not guaranteed
with many other mechanisms such as adaptive routing~\cite{Gratz2008,Fu2011},
deflective routing~\cite{Nilsson2003,Moscibroda2009,Ausavarungnirun2016}
or dynamic VC assignment~\cite{Nicopoulos2006}. Additionally, it is unnecessarily
restrictive for bypass conditions, since it prevents bypass in cases in which
it could be used.

The current paper reviews and extends the work originally announced in~\cite{Perez2018}.
This research proposes \emph{Non-Empty Buffer Bypass (NEBB)}, a novel approach for the use of the bypass that increases its utilization and effectiveness.
With \emph{NEBB}, bypassed packets can overtake buffered packets.
Our proposal can be implemented under different flow control mechanisms, and depending
on the buffer occupancy, it is more efficient to employ either
Wormhole (WH) or Virtual Cut-Through (VCT). Based on this observation, we
design \emph{NEBB-Hybrid}, a mechanism that dynamically selects between WH or VCT forwarding
in the bypass path, maximizing the amount of packets that use this shortcut.
The mechanisms have been implemented using private or shared buffers in mesh and torus topologies. 
In the later using Flit-Bubble Flow Control (FBFC~\cite{ma2015}) to efficiently avoid routing deadlock, given that NEBB does not require empty buffers to forward packets as required by other bypass router architectures~\cite{kumar2007,kumar2008,krishna2010}.
This paper extends and completes our previous seminal work with the following contributions:

\begin{itemize}
	\item An extension of the evaluations of~\cite{Perez2018} with a comparison between the flow control used in the original bypass routers in~\cite{kumar2007,kumar2008,krishna2010}, which requires empty buffers to forward packets, and the traditional WH mechanism used as our baseline.
    \item An implementation of NEBB in torus, using Flit-Bubble Flow Control (FBFC)~\cite{ma2015} with shared buffers.
    \item A detailed evaluation in mesh and torus NoCs, which shows reductions up to 75\% in buffer
        utilization that translate into latency and power reductions up to 30\% and 23\% respectively.
    \item An evaluation of NEBB under real-traffic workloads, using gem5~\cite{kumar2008} simulations driven by the PARSEC suite benchmarks~\cite{bienia2008parsec}.
    \item An analysis of potential starvation problems when giving priority to the bypass path over the standard path, which concludes that starvation is not an issue under realistic conditions.
\end{itemize}

The organization of the paper is as follows:
Section~\ref{sect:state_of_the_art} introduces lookahead bypass routers and flow control mechanisms;
Section~\ref{sect:hybrid_flow_control} describes the \emph{NEBB} and \emph{Hybrid} proposals, including the extension to torus networks with FBFC;
Section~\ref{sect:methodology} and~\ref{sect:evaluation} detail the methodology and present the results;
Section~\ref{sect:related} discusses relevant related work;
and Section~\ref{sect:conclusions} concludes the work.

\section{Background} \label{sect:state_of_the_art}
This section presents the required background on LookAhead bypass
routers and on the use of WH or VCT in the NoC.

\subsection{LookAhead Bypass Router Architecture}\label{subsect:lookahead_bypass_router_architecture}

\begin{figure}
    \centering
    \includegraphics[width=\columnwidth]{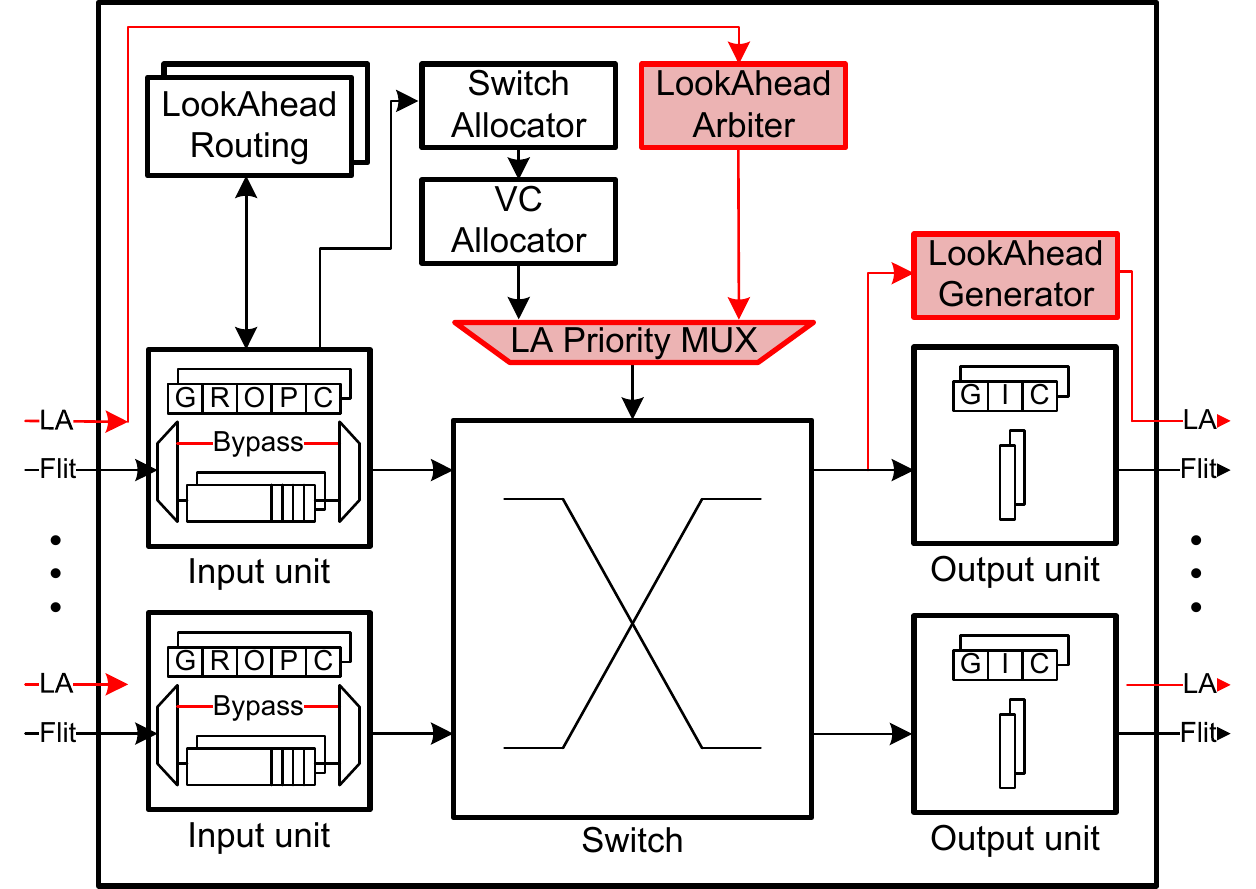}
    \caption{Lookahead (LA) bypass router architecture.}\label{fig:lookahead_bypass_router_architecture}
\end{figure}

This section describes a LookAhead bypass router architecture.
Lookahead bypass routers short-cut the buffer
write and arbitration pipeline stages in the absence of flit conflicts;
otherwise, the traditional (non-bypass) pipeline is used.  The implementation
relies on control packets, denoted \textit{advance bundles}, or
\textit{LookAheads} (LA), which setup the bypass one cycle before the arrival of
a flit.
LAs are generated after flits win access to the crossbar. They are destroyed
in the next router, after configuring the path or because of conflicts.

Figure~\ref{fig:lookahead_bypass_router_architecture} depicts the router
architecture based on the proposal of Krishna \emph{et al.}~\cite{krishna2010}.  The architecture
extends a traditional router with some additional units to support the bypass
forwarding (shaded in the figure): \emph{LookAhead Arbiter}, \emph{LA/flit
Priority} and \emph{LookAhead Generator}. The \emph{LookAhead Arbiter} is
optional as discussed in Section~\ref{sect:bypass_intro}; if not implemented, a
\emph{LookAhead Conflict Check} unit is required, which removes all
conflicting LAs. The \emph{LA/flit Priority} unit gives absolute priority to
the LAs over the buffered flits, or vice versa.

\begin{figure}
    \centering
    \includegraphics[width=\columnwidth]{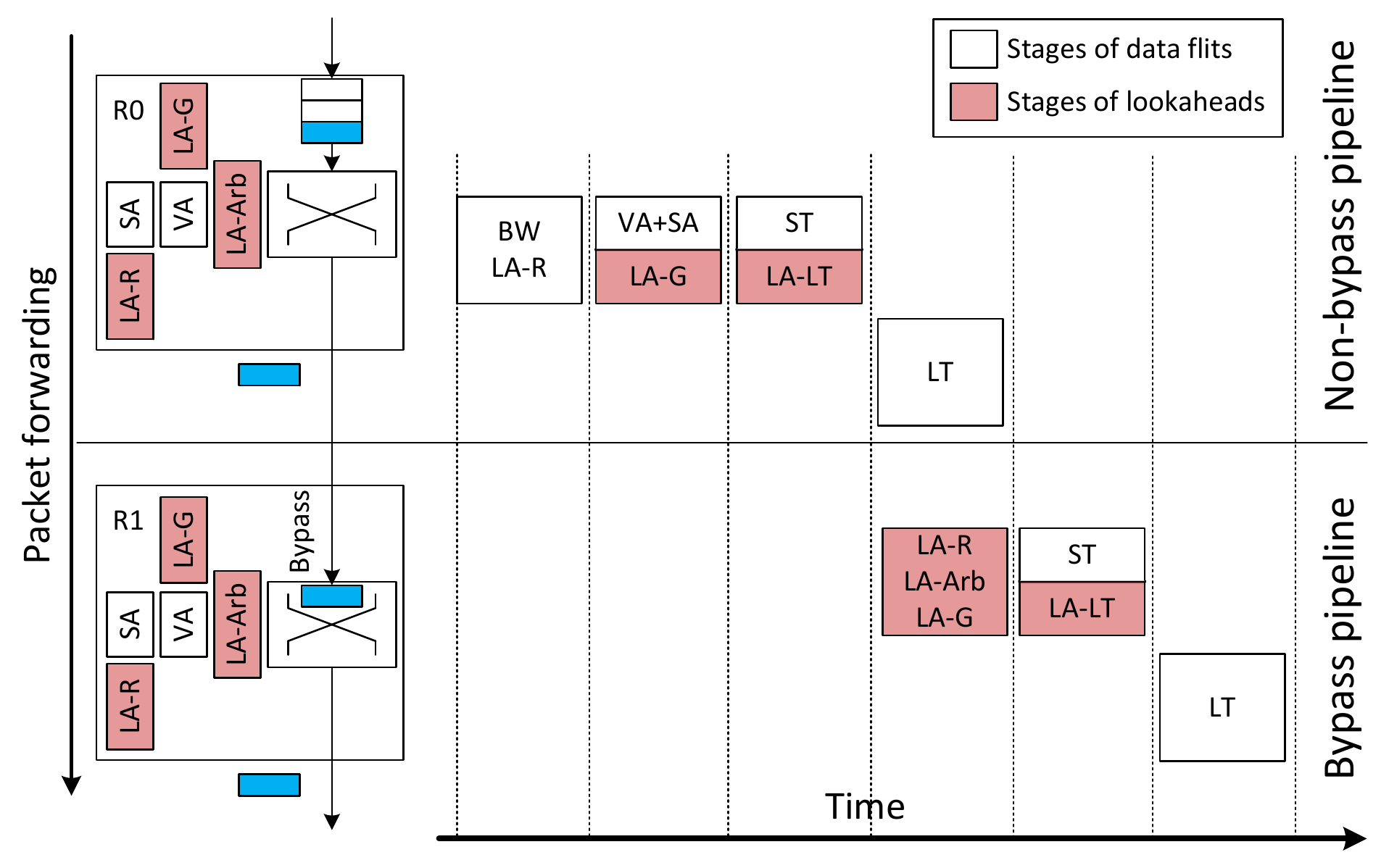}
    \caption{Lookahead bypass router pipelines. Acronyms are defined in the text in Section~\ref{subsect:lookahead_bypass_router_architecture}.}\label{fig:bypass_router_pipelines}
\end{figure}

Figure~\ref{fig:bypass_router_pipelines} represents the two pipelines of the
router, used in two consecutive routers R0 (non-bypass pipeline) and R1 (bypass
pipeline). White boxes
represent the pipeline stages for flits (data) and dark red boxes the stages of the
LAs used to pre-setup the crossbar. The functionality of each stage in both pipelines are
described next.

\subsubsection{Pipeline functions associated to flits}


\paragraph{Buffer Write (BW) and LA Routing (LA-R)}
Flits are received and stored in the
requested VC. When a head
flit reaches the front of a buffer, the LA route (for the next hop) is computed.

\paragraph{Virtual Channel Allocation (VA)} A head flit places a request for a
destination VC into the VC allocator. The VC allocator chooses winners for each
output among all input VCs. These allocations remain until a tail flit is
forwarded.

\paragraph{Switch Allocation (SA)} Flits are granted the use of the crossbar.
The switch allocator arbitrates among the output requests received from the
input ports in this stage.

\paragraph{Switch Traversal (ST) and Link Traversal (LT)} winners of SA
traverse the router crossbar to the output port, and then the link to
the next router or destination node.
When a LA wins access to the crossbar, the associated flit skips BW, VA and SA and goes to
the ST and LT stages.

The distribution of functions into stages may vary; Section~\ref{sect:methodology} presents the implementation used in our evaluations.

\subsubsection{Pipeline functions associated to lookaheads}


\paragraph{LookAhead Routing (LA-R)} it computes the packet route one hop in
advance, using the information from the received LA.

\paragraph{LookAhead Arbiter (LA-Arb)} LAs must reserve the desired output
port in order to set the bypass for the upcoming flit in the next cycle.
This functions handles conflicts between different LAs requesting the same
output port. Different implementations are considered: either no LA can
proceed in case of conflict (function LA-Conflict Check, no arbiter) or an arbiter
per output port is used to select one winner (function LA-Arb
as depicted in Figure~\ref{fig:lookahead_bypass_router_architecture}).

In addition, buffered flits in SA/VA stages can conflict with winning
lookaheads. In case of such conflicts, priority is given to either
buffered flits~\cite{kumar2007} or LAs~\cite{kumar2008, krishna2010}.

\paragraph{LookAhead Generation (LA-G) and Link Traversal (LA-LT)} it creates
the control information for a flit and sends it during its ST stage. In this way,
LAs arrive to the next router one cycle before their flits.

\subsection{Lookahead Bypass Router Policies}\label{sect:bypass_intro}


In a Lookahead bypass router, the bypass path is used only if the following \emph{bypass
conditions} are met:

\begin{enumerate}
    \item The buffer at the input port that receives the LA is empty.
    \item There is no output port conflict with buffered flits.
    \item There are no conflicts between LAs arriving in the same cycle.
\end{enumerate}

Condition 1 guarantees that packets do not interleave in the same buffer (as discussed
in section~\ref{sect:NEBB}) and are forwarded in order.
With multiple virtual channels, this restriction applies to the buffer of
the VC where the flit would be stored in case of using the non-bypass pipeline.
Many proposals employ a large number of VCs~\cite{kumar2007,kumar2008} to
avoid limitations in the bypass from this condition. However, this requires a
large buffer area and complicates the VC allocator, which typically sets the
critical path delay of the router~\cite{Jerger2017}.

Condition 2 gives absolute priority to flits in the non-bypass pipeline, this
is, those already stored in the pipeline buffers. Note that the opposite
priority may also be considered to maximize the utilization of the bypass path.
The impact of the selected priority is evaluated in
Section~\ref{sect:crossbar_priority_to_buffered_or_bypassed_flits}.

Condition 3 implies that there is no arbiter between LAs: when multiple
LAs contend for the same output, they are all discarded and the
associated flits use the non-bypass pipeline.

Different implementations may modify slightly these conditions. In~\cite{kumar2008, krishna2010}
absolute priority is given to LAs over packets in the buffers (condition 2 is
removed); and an arbiter per output port is implemented to select one wining LA in case of
conflict for an output port (condition 3).


\subsection{Flow Control and deadlock avoidance}\label{sect:WH_VCT}

Flow control in NoCs is applied at flit level when using WH. WH
forwards traffic on a flit-by-flit basis, based on the availability of space
for each flit in the buffer of the next router. WH allows to reduce buffer
size, since routers do not need to accommodate a complete packet.

However, to avoid unnecessary throttling buffers need to be sized for the
buffer turnaround time, this is, the minimum idle time required for
successive flits to reuse a buffer. A typical turnaround time is 4 to 6
cycles~\cite{Jerger2017,Becker2012}, similar to typical NoC packet sizes in
flits. Such amount of buffering allows to implement Virtual-Cut Through (VCT) flow
control, which forwards data on a packet-by-packet basis.

Multiple NoC designs rely on an additional restriction to forward data, requiring a free and empty VC in the next selected router to forward a packet. This simplifies the control, reducing the amount of signals sent between adjacent routers, and has been implemented in previous work such as ~\cite{kumar2007,kumar2008,krishna2010} 
Requiring the emptiness of the destination VC is a conservative solution to guarantee the first bypass condition mentioned in Section~\ref{sect:bypass_intro}. A packet that wins SA and starts to progress to the next router can only happen if the next VC is empty, hence LAs always find that the input buffer in the router to bypass is empty. These proposals follow a flit level switch allocation, so the employed flow control can be considered as a restricted implementation of WH. From now on, we refer to this flow control as \emph{WH - Empty VC} or simply \emph{Empty VC}.

Our proposal uses WH or VCT regulated by credits, without the previous emptiness requirement. This requires additional logic to check the first condition of the bypass. Every LA has to check, in the router to bypass, that the input buffer is empty and, in the case of WH, inactive as well. This is because in WH there can be holes (time gaps) between consecutive flits of a packet. Therefore, LAs can see the buffer empty but active, i.e. being used by another packet whose tail flit has not been sent yet.

\begin{figure}
	\centering
	\begin{subfigure}[b]{\columnwidth}
		\includegraphics[width=\textwidth]{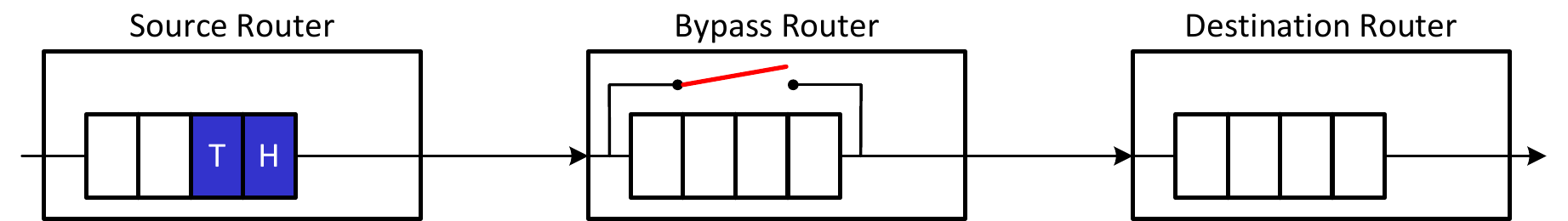}
		\caption{Initial buffer occupation to bypass the blue packet applying \emph{Empty VC}.}
		\label{fig:empty_vc_vs_wh_1}
	\end{subfigure}

	\begin{subfigure}[b]{\columnwidth}
		\includegraphics[width=\textwidth]{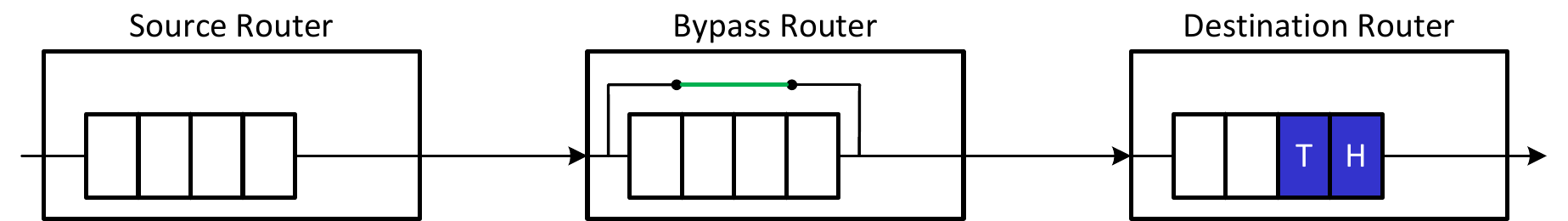}
		\caption{Bypass of the blue packet applying \emph{Empty VC}.}
		\label{fig:empty_vc_vs_wh_2}
	\end{subfigure}

	\begin{subfigure}[b]{\columnwidth}
		\includegraphics[width=\textwidth]{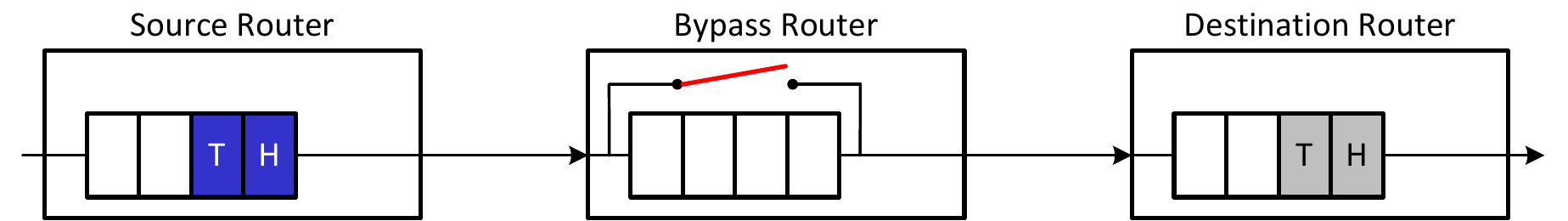}
		\caption{Initial buffer occupation to bypass the blue packet applying WH or VCT.}
		\label{fig:empty_vc_vs_wh_3}
	\end{subfigure}

	\begin{subfigure}[b]{\columnwidth}
		\includegraphics[width=\textwidth]{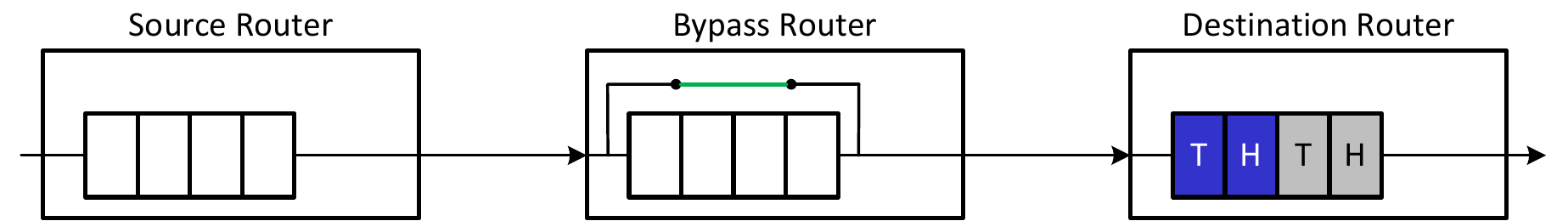}
		\caption{Bypass of the blue packet applying WH or VCT.}
		\label{fig:empty_vc_vs_wh_4}
	\end{subfigure}
	
	\caption{Example of packet bypassing applying \emph{Empty VC} and WH (VCT is equivalent in this example). \emph{Empty VC} requires both buffers empty, in the \emph{bypass router} and in the \emph{destination router}, to advance. WH or VCT requires space for a flit or packet, respectively to advance. Taking the bypass requires the emptiness of the \emph{bypass router} buffer and room for a flit (WH) or a packet (VCT) in the \emph{destination router}.}
	\label{fig:empty_vc_vs_wh}
\end{figure}

Figure~\ref{fig:empty_vc_vs_wh} depicts an example bypassing a packet (represented in blue), applying either \emph{Empty VC} (Figures~\ref{fig:empty_vc_vs_wh_1} and~\ref{fig:empty_vc_vs_wh_2}) or WH (Figures~\ref{fig:empty_vc_vs_wh_3} and~\ref{fig:empty_vc_vs_wh_4}).
There are three routers: \emph{source}, \emph{bypass} and \emph{destination}\footnote{They do not necessarily refer to the source and destination of the packet.}, with packets stored in their buffers.
\emph{Empty VC} requires the emptiness of the buffers in the \emph{bypass} and \emph{destination} routers. Nevertheless, our router implementations equipped with bypassing only requires the emptiness of the \emph{bypass router} buffer and room for a flit or a packet in the \emph{destination router} depending if the flow control applied is WH or VCT.

Many NoC designs employ link widths that accommodate a whole
packet~\cite{sodani2015knights,Kumar2017,Davidson2018}.  In such case, packets
are single-flit and there is no difference between WH and VCT.  Other designs
present bimodal traffic, with 1 and 5 flits being typical values~\cite{ma2015}.
For such cases, WH credit accounting is performed for each flit of the packet,
whereas in VCT it is done for the whole packet.
This accounting has to be carefully considered when both mechanisms are mixed,
particularly when using shared buffers.

Deadlock avoidance in a mesh is often implemented using routing restrictions, such as Dimension-Ordered Routing (DOR).
In torus networks additional restrictions are necessary. For example, Dateline~\cite{Dally2003} breaks the cyclic dependency chain of the physical links using at least 2 VCs. The packets in a ring initially only use one of the two VCs and the second VC is reserved for those packets that cross the Dateline, i.e. a predefined router in the ring. A more efficient solution is Flit-Bubble Flow Control (FBFC)~\cite{ma2015}, a derivative mechanism of Bubble Routing~\cite{Puente2001} for WH. These mechanisms leverage the flow control to avoid deadlock in the network. FBFC adds an injection rule to avoid deadlock: a packet is injected only when there is space for the whole packet plus one flit, i.e. the bubble. This rule guarantees the progress of flits that are already in the ring. The same applies to packets that change dimension.

\section{Efficient Bypass in Mesh and Torus} \label{sect:hybrid_flow_control}
This section first introduces our Non-Empty Buffer Bypass proposal, detailing
VCT and WH implementations. Then, it introduces the \emph{Hybrid} mechanism
which combines VCT and WH to maximize bypass. Finally, some implementation
details are presented.

\subsection{Non-Empty Buffer Bypass}\label{sect:NEBB}

Non-Empty Buffer Bypass (NEBB) performs bypass
when the router buffer is not empty, as long as this does not interleave
flits from different packets in a single virtual channel.
NEBB relaxes the bypass condition 1 introduced in
Section~\ref{sect:bypass_intro} to the following two general conditions, which depend
on the flow control mechanism:
\begin{enumerate}[label=1\alph*)]
    \item No packet in the bypassed input VC is already advancing to an output port.
    \item The packet may be forwarded without packet interleaving in a buffer, according to the flow control employed.
\end{enumerate}

The VC control registers (GROPC in Figure~\ref{fig:lookahead_bypass_router_architecture}) are
updated for both bypass and ordinary packet forwarding. Condition~1a is required to avoid
that bypassed packets conflict with other packets that have already won allocation and
have their status recorded in these VC control registers.

\begin{figure}[t]
    \centering

    \begin{subfigure}[b]{\columnwidth}
        \includegraphics[width=\textwidth]{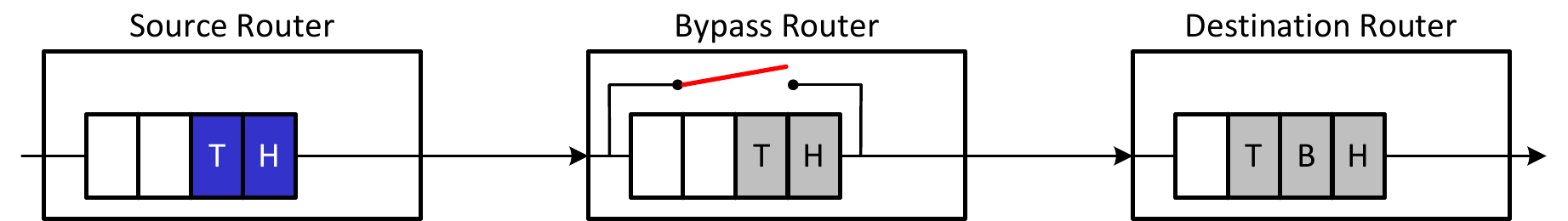}
        \caption{Initial state. $H$, $B$ and $T$ denote \textit{head},
            \textit{body} and \textit{tail} flit.}
        \label{fig:case0-initial_state}
    \end{subfigure}

    \begin{subfigure}[b]{\columnwidth}
        \includegraphics[width=\textwidth]{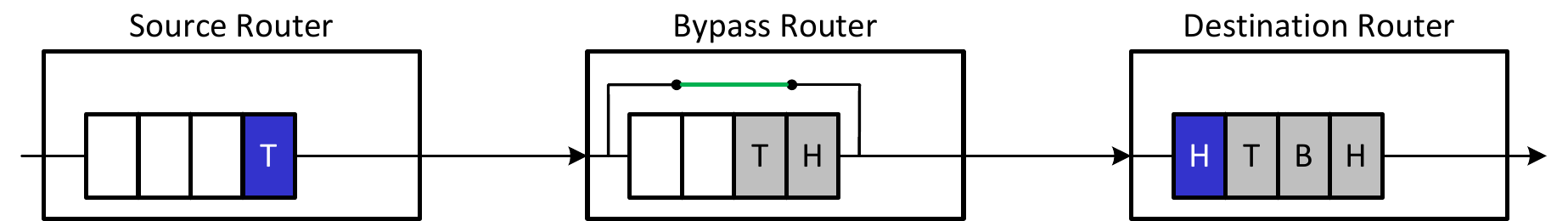}
        \caption{Incorrect bypass of the blue head flit.}
        \label{fig:case0-head_bypass}
    \end{subfigure}

    \begin{subfigure}[b]{\columnwidth}
        \includegraphics[width=\textwidth]{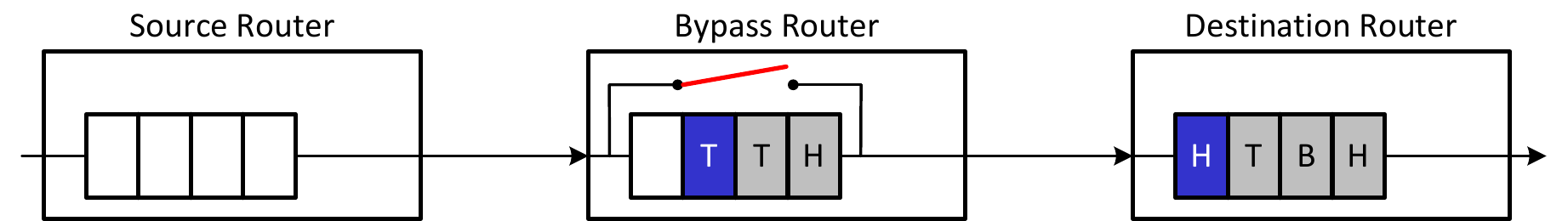}
        \caption{Blue tail flit is buffered at the \emph{bypass router} because there is no room
            at the \emph{destination router}, causing packet interleaving in the same buffer and data
            corruption.}
        \label{fig:case0-interleave}
    \end{subfigure}

    \caption{Incorrect packet bypass, which interleaves data in the same buffers.}
    \label{fig:case0}
\end{figure}

Condition~1b is required to prevent data corruption, and it is dependent on
the flow control used. Figure~\ref{fig:case0} shows an example of \emph{incorrect} packet bypass
using WH, to illustrate this requirement.
The dark blue packet in the \emph{source} router in Figure~\ref{fig:case0-initial_state}
tries to bypass the intermediate router. The Head (H) flit is bypassed in Figure~\ref{fig:case0-head_bypass}
because there is a free buffer in the \emph{destination} router buffer (following
WH) and overtakes the packet in the non-empty buffer of the \emph{bypass}
router. The next flit (tail, T) cannot be bypassed because there is no more room
in the \emph{destination router} buffer, so it is stored in the buffer of the
\emph{bypass router} in Figure~\ref{fig:case0-interleave}, behind the existing packet.
In this situation, the grey packet is interleaved in the same VC with the blue packet,
so data is corrupted. Even if the grey packet is forwarded to a different output,
the blue tail flit has lost its routing and status information from the VC registers,
so it cannot be forwarded.

Table~\ref{tab:rules} summarizes the different cases in which bypass can be applied for
each flow control mechanism, with and without \emph{NEBB}. Note that in the base mechanisms without \emph{NEBB}
both implementations of WH, with and without the \emph{Empty VC} requirement, present the same behavior because
the requirement of the empty buffer is already included in the original bypass conditions. 
The specific forwarding conditions for \emph{NEBB} using WH or VCT are presented next.
These conditions rely on the occupancy level of the buffers
in the \emph{bypass} and \emph{destination} routers. 

\begin{table*}
    \caption{Bypass buffer conditions for different mechanisms. \emph{bypass buffer} refers to the buffer in the bypass router being empty or not. \emph{Dest. buffer} indicates if the destination buffer may accommodate the whole \emph{packet} or only some flits (\emph{partial})}
    \label{tab:rules}
    \centering
    \resizebox{\textwidth}{!}{
        \begin{tabu}{|[1.5pt]l|c|c|c|c|c|c|} \tabucline[1.5pt]{-}
        	
            \multirow{3}{0.15\textwidth}{\textbf{Bypass type (Required buffer size)}} & \multicolumn{3}{|[1.5pt]c|}{\textbf{Bypass buffer: empty}} & \multicolumn{3}{|[1.5pt]c|[1.5pt]}{\textbf{Bypass buffer: not empty}} \\ \cline{2-7}
             & \multicolumn{1}{|[1.5pt]c|}{\textbf{Single-flit packet}} & \multicolumn{2}{c|}{\textbf{Multi-flit packet}}                                                                & \multicolumn{1}{|[1.5pt]c|}{\textbf{Single-flit packet}}  & \multicolumn{2}{c|[1.5pt]}{\textbf{Multi-flit packet}}  \\\cline{2-7}
             & \multicolumn{1}{|[1.5pt]c|}{}                            & \textbf{Dest. buffer: packet}  & \textbf{Dest. buffer: partial}    & \multicolumn{1}{|[1.5pt]c|}{}                             & \textbf{Dest. buffer: packet} & \multicolumn{1}{c|[1.5pt]}{\textbf{Dest. buffer: partial}} \\\tabucline[1.5pt]{-}
            VCT (packet size)      & \multicolumn{1}{|[1.5pt]c|}{{\cellcolor{green!5}}\true} & {\cellcolor{green!5}}\true & {\cellcolor{red!5}}\false & \multicolumn{1}{|[1.5pt]c|}{{\cellcolor{red!5}}\false} & \multicolumn{1}{|c|}{{\cellcolor{red!5}}\false} & \multicolumn{1}{|c|[1.5pt]}{{\cellcolor{red!5}}\false} \\\hline
            WH - Empty VC\footnotemark[2] (1 flit) & \multicolumn{1}{|[1.5pt]c|}{{\cellcolor{green!5}}\true} & {\cellcolor{green!5}}\true & {\cellcolor{green!5}}\true & \multicolumn{1}{|[1.5pt]c|}{{\cellcolor{red!5}}\false} & \multicolumn{1}{|c|}{{\cellcolor{red!5}}\false} & \multicolumn{1}{|c|[1.5pt]}{{\cellcolor{red!5}}\false} \\\hline
            WH - Baseline (1 flit) & \multicolumn{1}{|[1.5pt]c|}{{\cellcolor{green!5}}\true} & {\cellcolor{green!5}}\true & {\cellcolor{green!5}}\true & \multicolumn{1}{|[1.5pt]c|}{{\cellcolor{red!5}}\false} & \multicolumn{1}{|c|}{{\cellcolor{red!5}}\false} & \multicolumn{1}{|c|[1.5pt]}{{\cellcolor{red!5}}\false} \\\tabucline[1.5pt]{-}
            NEBB-WH (1 flit) & \multicolumn{1}{|[1.5pt]c|}{{\cellcolor{green!5}}\true} & {\cellcolor{green!5}}\true & {\cellcolor{green!5}}\true & \multicolumn{1}{|[1.5pt]c|}{{\cellcolor{green!5}}\true} & \multicolumn{1}{|c|}{{\cellcolor{red!5}}\false} & \multicolumn{1}{|c|[1.5pt]}{{\cellcolor{red!5}}\false} \\\hline
            NEBB-VCT (packet size) & \multicolumn{1}{|[1.5pt]c|}{{\cellcolor{green!5}}\true} & {\cellcolor{green!5}}\true & {\cellcolor{red!5}}\false & \multicolumn{1}{|[1.5pt]c|}{{\cellcolor{green!5}}\true} & \multicolumn{1}{|c|}{{\cellcolor{green!5}}\true} & \multicolumn{1}{|c|[1.5pt]}{{\cellcolor{red!5}}\false}\\\tabucline[1.5pt]{-}
            NEBB-Hybrid (1 flit\footnotemark[3]) & \multicolumn{1}{|[1.5pt]c|}{{\cellcolor{green!5}}\true} & {\cellcolor{green!5}}\true & {\cellcolor{green!5}}\true & \multicolumn{1}{|[1.5pt]c|}{{\cellcolor{green!5}}\true} & \multicolumn{1}{|c|}{{\cellcolor{green!5}}\true} & \multicolumn{1}{|c|[1.5pt]}{{\cellcolor{red!5}}\false} \\\tabucline[1.5pt]{-}
        \end{tabu}
    }
\end{table*}

\subsubsection{NEBB-WH}\label{sect:NEBB-WH}

Under wormhole, arbitration is performed flit by flit. Therefore, the bypass of
one flit does not guarantee that the following flits of the packet will be
also bypassed. Figure~\ref{fig:case0} has already illustrated an incorrect
case of using WH with NEBB. For this reason,
multi-flit packet bypass is unsafe when the \emph{bypass router} buffer is not empty.

The solution for NEBB-WH is to avoid bypassing multi-flit packets when the \emph{bypass router}
buffer is not empty. By contrast, NEBB-WH can bypass single-flits packets, which is a frequent
case. Therefore, condition 1 under WH flow control results as follows:

\begin{enumerate}[label=1\alph*)]
    \item No packet in the input buffer (VC) is already advancing to an output port.
    \item The packet is single-flit or the bypass buffer is empty.
\end{enumerate}

\subsubsection{NEBB-VCT}\label{sect:NEBB-VCT}

Under Virtual Cut-Through, arbitration is performed once per packet and the assigned
resources remain allocated for the duration of the packet forwarding. This prevents any
packet interleaving, avoiding the problem presented in Figure~\ref{fig:case0}.

However, to forward a packet VCT requires space for the whole packet at the destination
buffer. This means that in NEBB-VCT multi-flit packets cannot be bypassed if the
\emph{destination router} buffer can only accommodate part of the packet. Additionally,
the buffer in the \emph{bypass router} also needs room to accommodate the whole packet,
even if it is not used: otherwise, the \emph{source router} would not start sending
the packet. Therefore, condition 1 under VCT results as follows:

\begin{enumerate}[label=1\alph*)]
    \item No packet in the input buffer (VC) is already advancing to an output port.
    \item The bypass and destination buffers have room for the whole packet.
\end{enumerate}

\begin{figure}
    \centering
    \begin{subfigure}[b]{\columnwidth}
        \includegraphics[width=\textwidth]{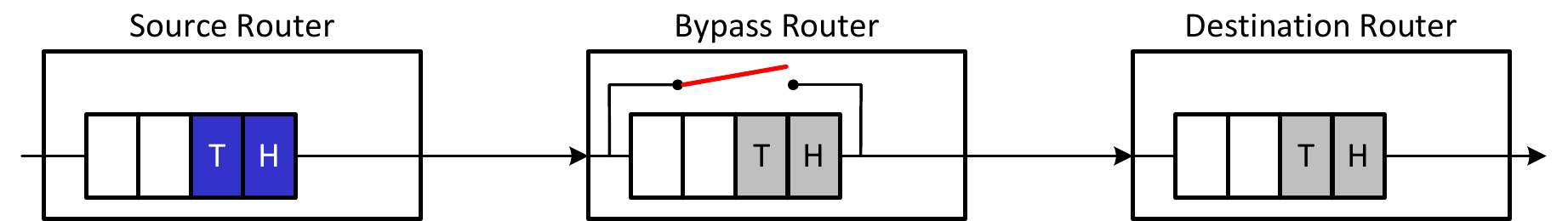}
        \caption{Initial state of the example.}
        \label{fig:case1-init}
    \end{subfigure}

    \begin{subfigure}[b]{\columnwidth}
        \includegraphics[width=\textwidth]{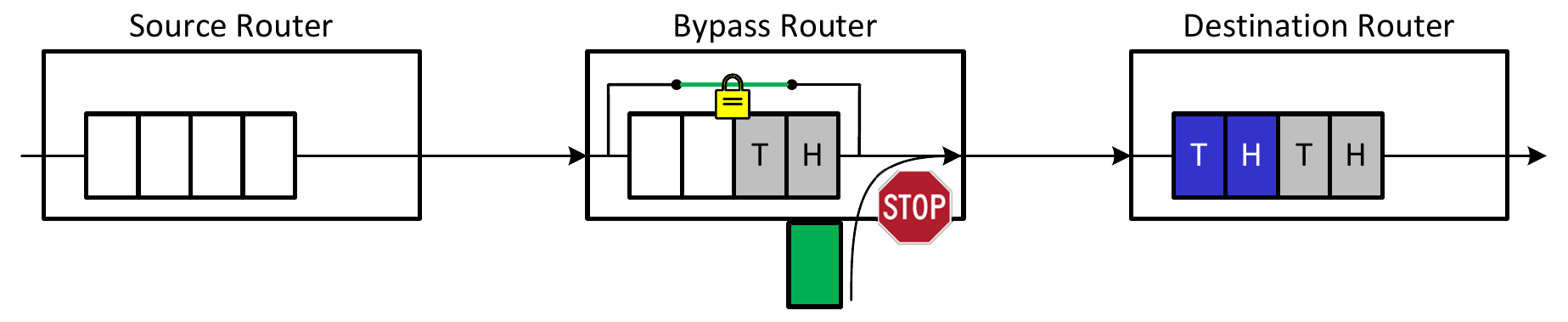}
        \caption{Bypass allowed as there is room for the blue packet in the \emph{destination}.}
        \label{fig:case1-hybrid}
    \end{subfigure}

    \caption{NEBB VCT bypass. A packet is bypassed to the \emph{destination router}, with room for the whole
    packet in the \emph{destination} and a not empty buffer in the \emph{bypass router}.}
    \label{fig:case1}
\end{figure}

Figure~\ref{fig:case1-hybrid} illustrates the bypass of packets in VCT.
As there is room for the whole packet in the \emph{bypass} and \emph{destination routers},
the blue packet can be bypassed, independently of the emptiness of the
buffer in the \emph{bypass router}. During packet bypass the resources are reserved (locked),
so other buffered flits or lookaheads (e.g. the green flit in a different port) cannot
obtain the output port that is using the blue packet, until the tail flit of the blue
packet reaches the ST stage of the \emph{bypass router}.

\subsection{Improved bypass in NEBB via \emph{Hybrid} Flow Control}\label{sect:hybrid}

The \emph{NEBB-Hybrid} mechanism (or simply \emph{Hybrid}) dynamically selects between WH and VCT
to maximize the utilization of the bypass. Section~\ref{sect:NEBB} and Table~\ref{tab:rules} present the
limitations of \emph{NEBB} when using each mechanism: \emph{NEBB-WH} does not bypass
multi-flit packets when the \emph{bypass router} buffer is not empty;
\emph{NEBB-VCT} does not bypass multi-flit packets when the \emph{bypass router}
buffer cannot accommodate the whole packet. \emph{Hybrid} selects the most suitable
mechanism in each case.

In \emph{Hybrid} the standard pipeline uses WH. The bypass pipeline uses both
flow controls: if the buffer to bypass is empty, the router checks if there is room for a flit in the
destination VC, following WH; otherwise, the router
checks if there is room for the whole packet, following VCT. For single-flit packets,
both mechanisms are equivalent.

Combining two different flow control mechanisms introduces subtle implementation requirements.
The \emph{Hybrid} system employs WH, so
packets from different VCs may be interleaved in the same physical channel.
In such case, a ``hole'' appears in the forwarding of the flits of a packet.
Allocation in \emph{Hybrid} is implemented flit by flit, to take advantage of such holes
and maximize forwarding.

However, the original NEBB-VCT forwarding for multi-flit packets is safe because it guarantees that
the whole packet is forwarded consecutively, as presented in Section~\ref{sect:NEBB-VCT}. To preserve this
safeguard with flit by flit allocation and VCT, \emph{Hybrid} employs variable priority arbiters in
the LA Arbiter. Maximum priority is assigned to the LAs of multi-flit packets bypassed by VCT.
For each output port, there can be at most one packet with maximum priority. The holes of
this packet are leveraged to bypass flits from other packets, but only following WH (including
single-flit packets), to avoid having to assign maximum priority to two packets in the same
output port.

\subsection{Credit management using shared buffers}\label{sect:shared_buffers_credit_management}\label{sect:implementation_details}



Shared buffers~\cite{Tamir1992,Nicopoulos2006} improve efficiency.
Shared buffer capacity accounting needs to consider the dual flow control.
Packets advancing following VCT have to reserve room for their size in advance. Otherwise,
other packet advancing to another VC of the same input (using WH)
may invade slots initially intended for the first packet.

The evaluations of this work use credits, so the reservation is done
decrementing the credit count by the packet size when bypassing a packet
via VCT, or flit by flit via WH.

\footnotetext[2]{Requires the emptiness of the destination buffer.}
\footnotetext[3]{Packets larger than the buffer size cannot be bypassed by VCT rules.}

\subsection{VC management: hardware considerations}

The router architectures proposed in~\cite{kumar2007,krishna2010} combine a pool of free VCs together with on/off signaling and a VC Selector (VS), instead of traditional Virtual channel Allocation (VA), to simplify the VC management and, therefore, reduce the length of the input path that may determine the critical path. The \emph{on/off} signalings or \emph{free VC} signals, replace credits by indicating to the upstream router whether the current router has a free VC in the pool or not. VS selects a free VC when a packet arrives to the router instead of assigning the VC in the upstream router after a packet wins VA.
The proposed router architecture for the different versions of NEBB uses traditional credits and VA, but it is compatible with on/off signaling and VS. Instead of using \emph{free VC} signals, NEBB can use \emph{avail VCs} signals to indicate if there is an available VC with enough available slots to store a whole packet. In the case of NEBB-VCT and NEBB-Hybrid, they require one \emph{avail VC} signal per packet size, in order to perform VCT forwarding when bypassing the whole packet.

Another implication of implementing NEBB is the extra logic necessary to check the additional bypass conditions. We assume that the overhead incurred is low because the conditions are simple and they can be pre-evaluated in the previous cycle. Besides, NEBB does not require VCs to achieve a high bypass utilization and throughput (see Section~\ref{sect:evaluation}) so the complexity of the VC management can be reduced drastically if required to meet timing requirements.

\subsection{\emph{NEBB} in torus using Flit Bubble Flow Control} \label{sect:switch_allocator}

Tori are considered good alternatives to meshes~\cite{shin2011,singh2003goal} because they are node-symmetric, have lower diameter and double the bisection bandwidth of meshes due to their wraparound links. Like 2D-meshes, 2D-tori adapt well to the tiled organization of CMPs. However, wraparound links introduce routing cyclic dependencies.

This subsection presents the implementation details required to adapt \emph{NEBB} to a torus topology. In particular, the proposed design supports a torus using FBFC for deadlock avoidance and shared buffers, which is an efficient implementation. FBFC is introduced first. It is followed by a discussion on the implementation details of the allocation mechanisms, which may cause deadlock in torus topologies, and the specific details to avoid it when using \emph{NEBB}.

Flit Bubble Flow Control (FBFC)~\cite{ma2015} is an efficient deadlock avoidance mechanism
for torus-like topologies.
FBFC-L (Localized) requires that when a packet is injected or changes its
dimension there must be space for the whole packet plus one flit.
This guarantees that at least one flit slot is always empty in a dimension.
Note that FBFC does not support the \emph{Empty VC} mechanism described in Section~\ref{sect:WH_VCT}. In \emph{Empty VC} only one packet can coexist in a VC. A flit from another packet cannot advance in the ring and occupy the bubble because there is another packet (the injected one) in the buffer.

The switch allocator often employs a two-stage implementation~\cite{kumar2008,krishna2010},
first an arbiter for the inputs and then an arbiter for each output. Simple round-robin (RR)
arbiters are often used. RR input arbiters cycle through all available VCs, selecting one at
a time consecutively. If one of the output VCs is not available (for example, there are no credits in
the destination VC) it does not proceed, wasting one cycle. However, such implementation
simplifies the router design, since output availability does not need to be propagated to the
inputs. Also, such implementation inherently multiplexes packet flits, generating packet holes
in WH.

Holes are undesirable when VCT is also employed in the bypass in \emph{Hybrid}, as discussed
in Section~\ref{sect:hybrid}. To minimize them, we use a variable priority input arbiter, selecting
the same VC until the packet tail flit is forwarded, similar to VCT. However, this might introduce
performance and deadlock issues when WH and VCT are combined: First, a packet may be forwarded
by WH without space at the destination buffer for the complete packet, introducing delays until
the buffer becomes available. Second, this introduces a dependence between the input VCs,
which generates a deadlock when a delayed packet in a buffer waits for another buffer that is full
and waiting for the local arbiter to finish forwarding a packet from a different VC, also delayed. 
This deadlock problem is explained in more detail next with an example.

\begin{figure}[t]
	\centering
	
	\begin{subfigure}[b]{\columnwidth}
		\includegraphics[width=\textwidth]{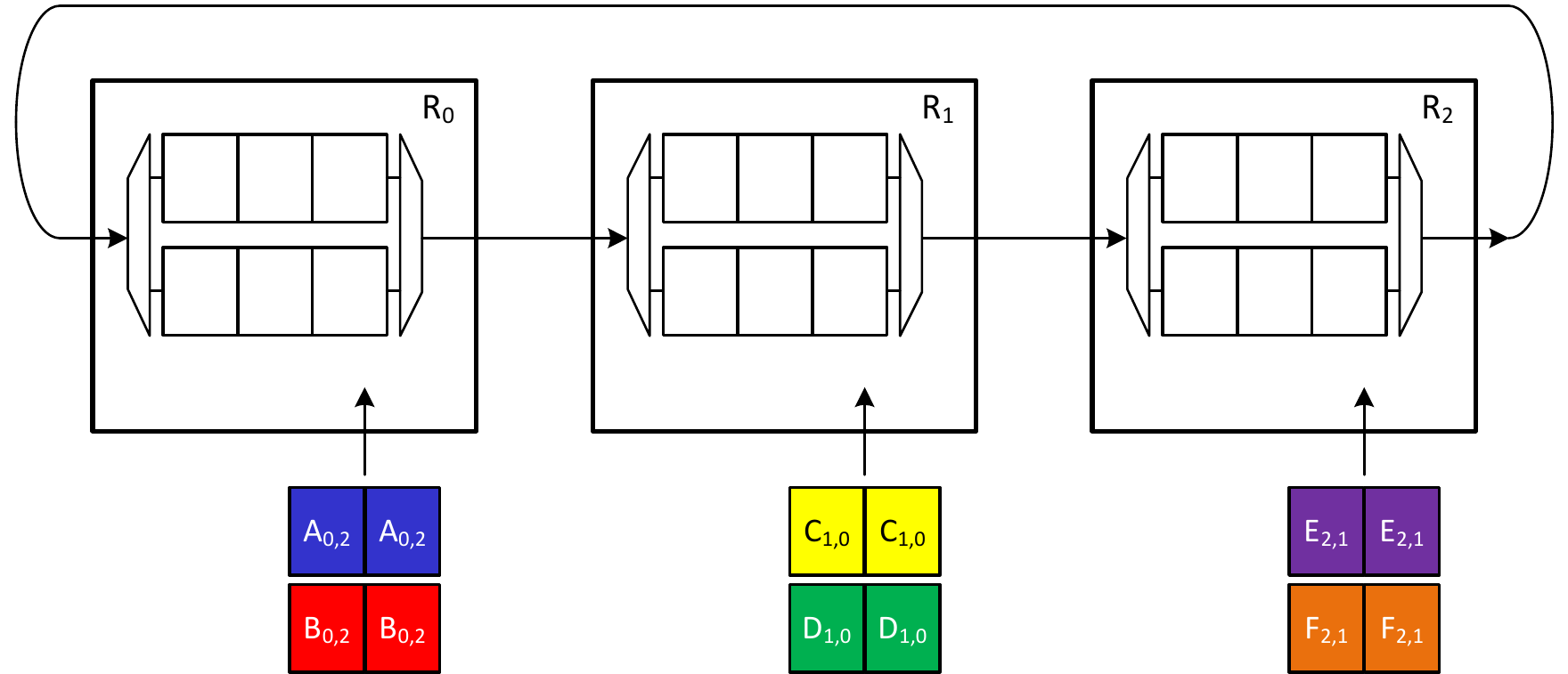}
		\caption{Initially each router injects two packets.}
		\label{fig:fbfcl_deadlock_1}
	\end{subfigure}

	\begin{subfigure}[b]{\columnwidth}
		\includegraphics[width=\textwidth]{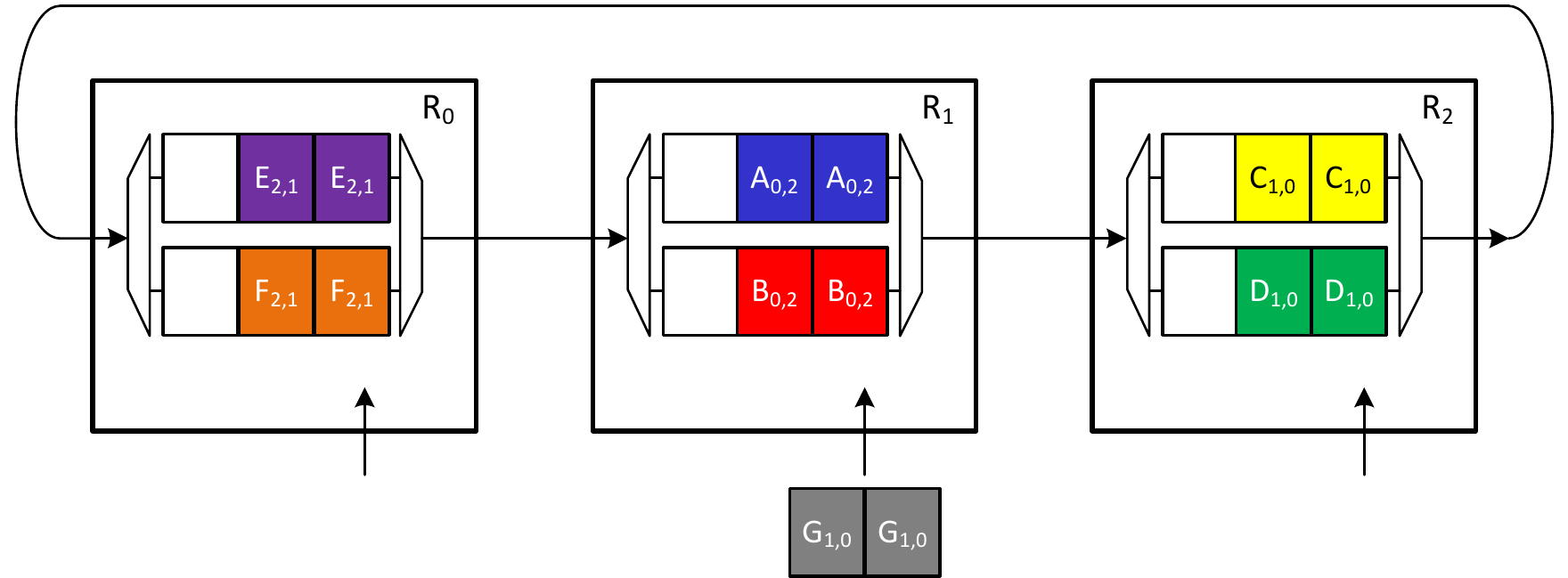}
		\caption{$R_1$ injects packet $G$ after sending packet $D$ to $R_0$.}
		\label{fig:fbfcl_deadlock_2}
	\end{subfigure}
	
	\begin{subfigure}[b]{\columnwidth}
		\includegraphics[width=\textwidth]{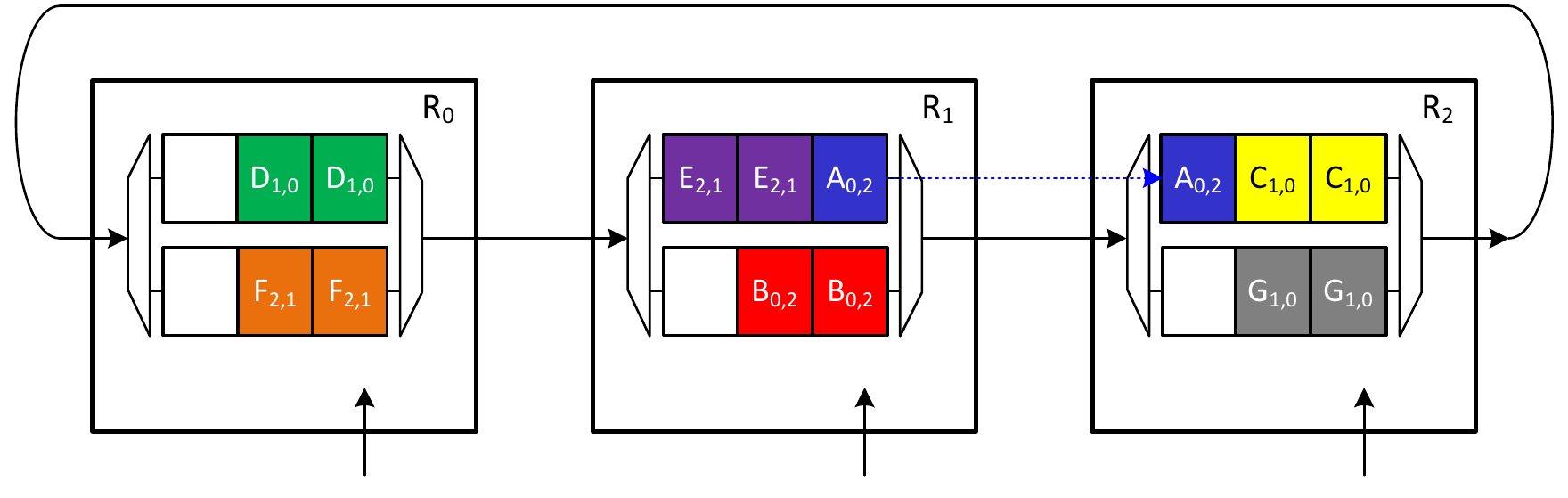}
		\caption{SA-I winners: $E$ in $R_0$, $A$ in $R_1$ and $D$ in $R_2$.}
		\label{fig:fbfcl_deadlock_3}
	\end{subfigure}
	
	\begin{subfigure}[b]{\columnwidth}
		\includegraphics[width=\textwidth]{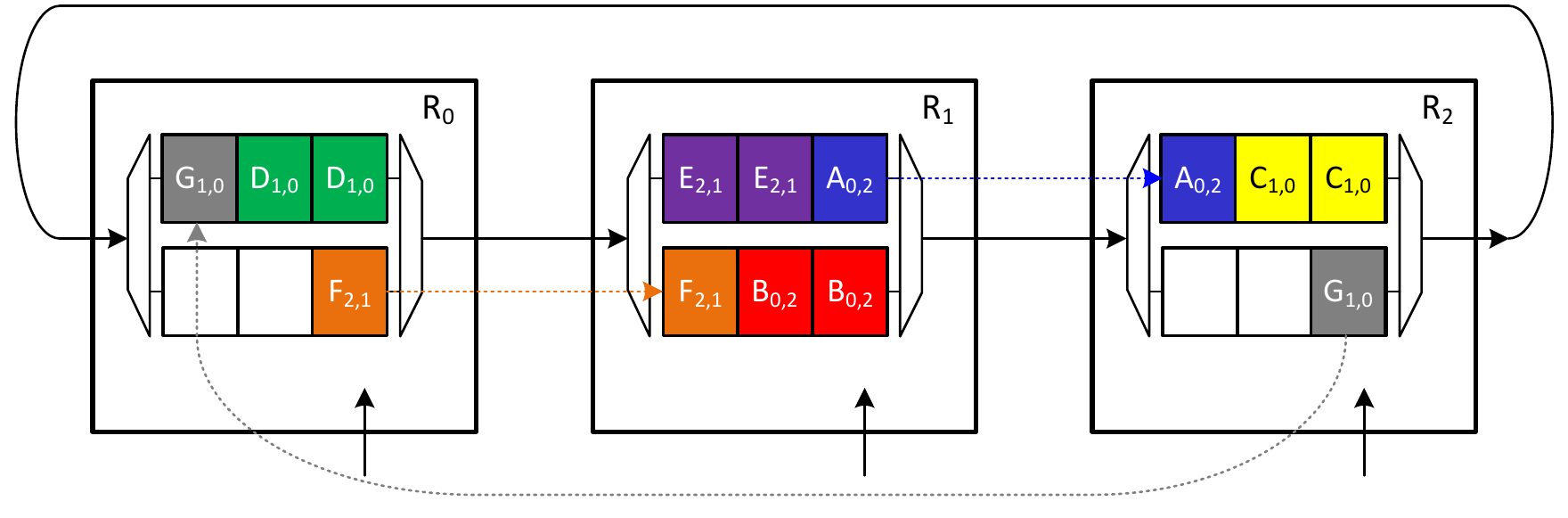}
		\caption{$F$ wins in $R_0$ and $F$ in $R_2$ producing a deadlock.}
		\label{fig:fbfcl_deadlock_4}
	\end{subfigure}

	\caption{Switch allocator deadlock in Torus (ring) using FBFC. Packets have two flits. They are represented with different colours and letters ($X_{src, dst}$, where $X$ the packet identifier, $src$ the source router and $dst$ the destination router).}
	\label{fig:fbfcl_deadlock}
\end{figure}

The configuration presented in Figure~\ref{fig:fbfcl_deadlock} has three routers forming a ring (1D-Torus), unidirectional in this case for simplicity. Every packet in the example has two flits and requires two hops to reach the destination router. Each packet is represented with a different colour and is identified by a letter, its source and destination: $X_{src,dst}$, where $X$ is the identifier, $src$ the source and $dst$ the destination.

Figure~\ref{fig:fbfcl_deadlock_1} depicts the initial state, where two packets are injected in each router: $A$ and $B$ in $R_0$; $C$ and $D$ in $R_1$; and $E$ and $F$ in $R_2$. The FBFCL condition is satisfied for every packet because, at least, there is a VC in the next router with room for 3 flits (2 flits plus a bubble).

Figure~\ref{fig:fbfcl_deadlock_2} represents the state of the routers after injecting these packets in the ring. In this situation there is a conflict between VCs in SA-I of each router. Assume that the winners are: $E$ in $R_0$, $A$ in $R_1$ and $D$ in $R_2$. And that they choose the first VC of $R_1$, $R_2$ and $R_0$, respectively.
In addition, $R_1$ will inject a new packet, $G$, in the ring after $D$ frees the second VC of $R_2$.
The new state of the routers is depicted in Figure~\ref{fig:fbfcl_deadlock_3}. In this case $F$ wins in $R_0$ and $G$ in $R_2$. $G$ chooses the first VC of $R_0$ and $F$ the second VC of $R_1$.
Finally, Figure~\ref{fig:fbfcl_deadlock_4} shows the deadlock. In $R_0$, $F$ locks SA-I but cannot progress because the second VC of $R_1$ is full. The same occurs with $A$ in $R_1$. $B$ cannot advance to $R_2$ despite there is space the second VC because SA-I is locked by $A$. And in $R_2$, $G$ is blocked by the lack of space in the first VC of $R_0$, and $C$ cannot win SA-I because it is locked by $G$.

To avoid these issues, we give priority to body flits to minimize packet holes, but remove
this priority when a flit does not advance, so the following ready VC is selected in the next
arbitration cycle.

\subsubsection{Bypass in torus using Flit Bubble Flow Control and shared buffers}


The bypass mechanisms are compatible with a torus using FBFC and private buffers.
However, shared buffers may introduce deadlock because the bubble condition is checked when the
head of the packet is forwarded. Consider a shared buffer with space for exactly two packets.
Two packets are forwarded, interleaved in the same physical channel, and they obey the condition
when their head is sent. However, when they are fully received, the bubble disappears.

To avoid this issue, when multi-flit packets are forwarded to a shared buffer with
a dimension change, the credit counter is decremented by the whole packet size when the
head flit is sent, regardless of the flow control. Therefore, in \emph{NEBB-Hybrid}
credits are decremented by the whole packet in two cases: when multi-flit packets
are forwarded using VCT and for injection and dimension change using any flow control.

\section{Methodology} \label{sect:methodology}
\begin{figure*}[t]
	\centering
	\begin{subfigure}[b]{\textwidth}
		\includegraphics[width=\textwidth]{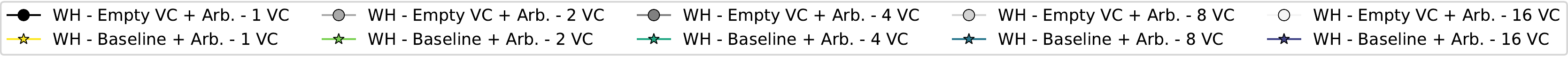}
	\end{subfigure}
	
	\subcaptionbox{DAMQ. Total size: 10 flits.\label{fig:on_off_shared_buf_size_10}}{
		\includegraphics[height=9\baselineskip]{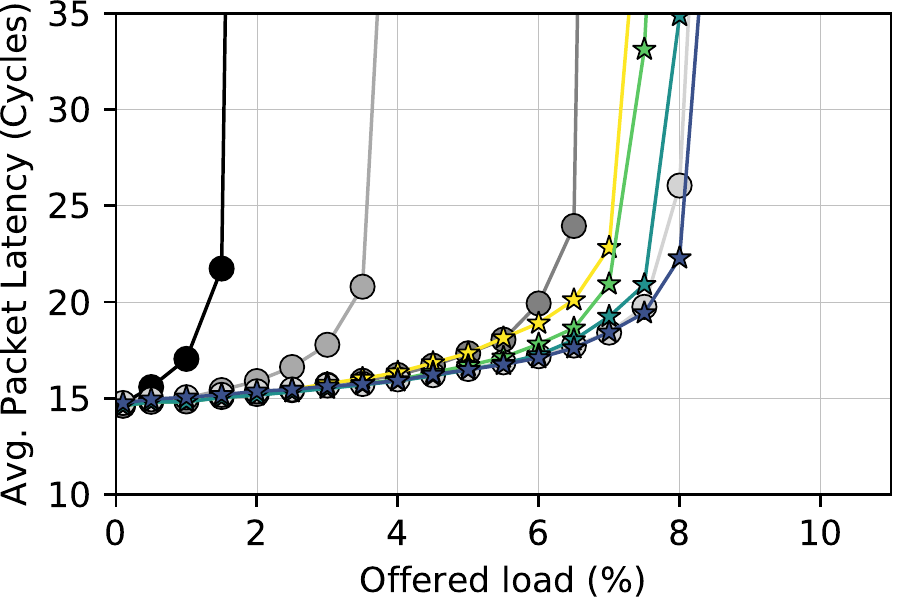}}\hspace{1cm}
	\subcaptionbox{DAMQ. Total size: 20 flits.\label{fig:on_off_shared_buf_size_20}}{
		\includegraphics[height=9\baselineskip]{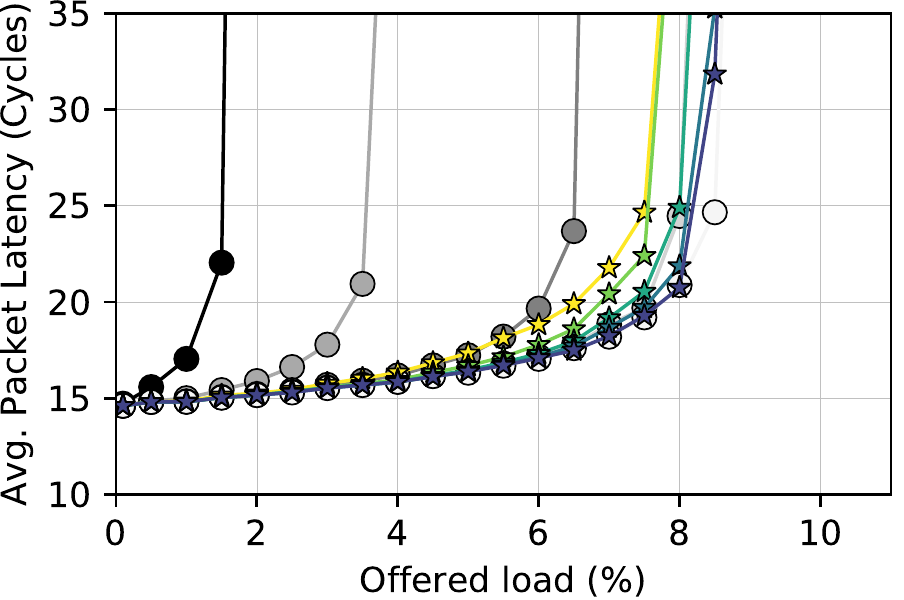}}
	
	\subcaptionbox{Private VC. VC size: 5 flits.\label{fig:on_off_private_buf_size_5}}{
		\includegraphics[height=9\baselineskip]{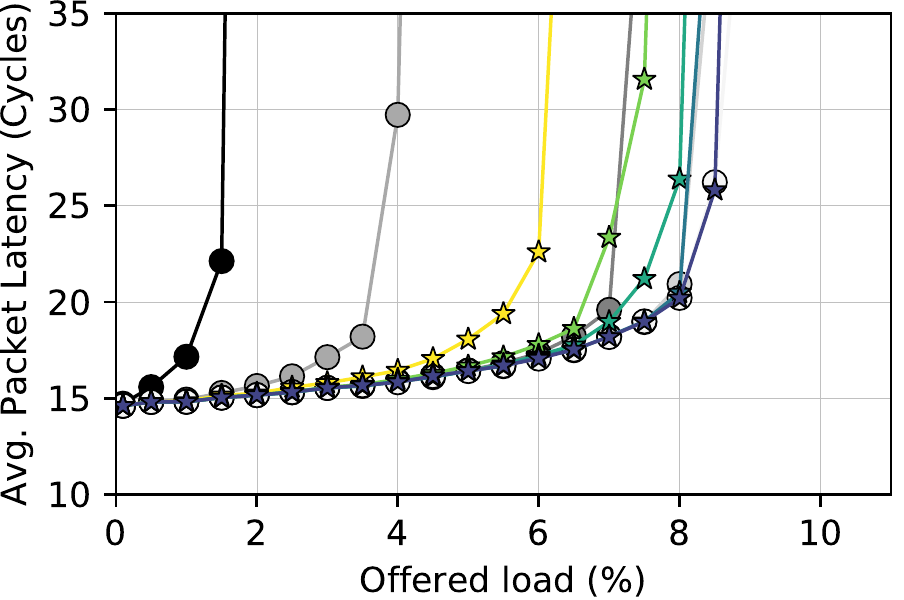}}\hspace{1cm}
	\subcaptionbox{Private VC. VC size: 10 flits.\label{fig:on_off_private_buf_size_10}}{
		\includegraphics[height=9\baselineskip]{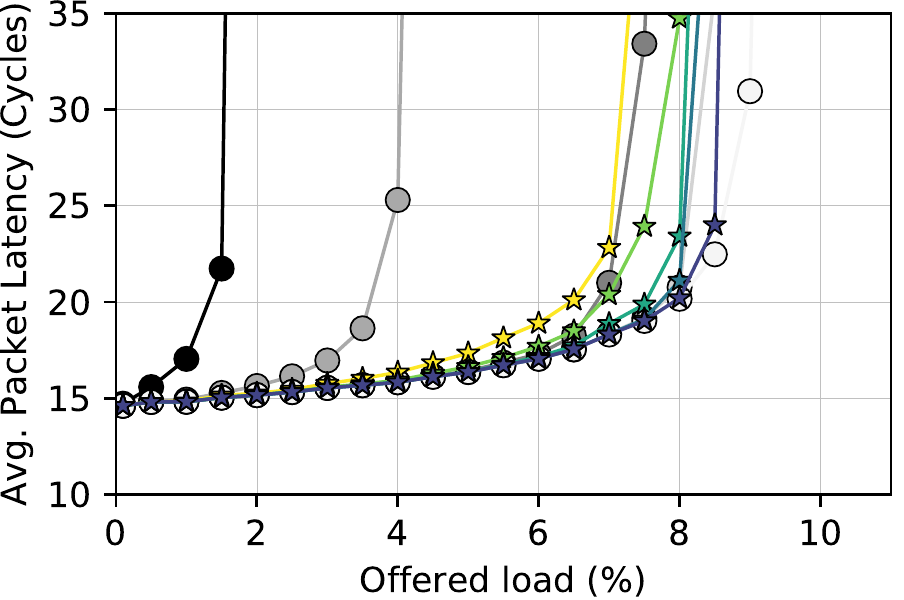}}
	
	\caption{Packet latency in an $8 \times 8$ $c=4$ mesh with \emph{Empty VC} and WH flow controls, using bimodal traffic.}\label{fig:on_off}
\end{figure*}

\begin{figure*}[t]
	\centering
	\begin{subfigure}[b]{0.38\textwidth}
		\includegraphics[width=\textwidth]{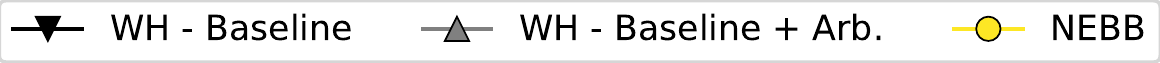}
	\end{subfigure}
	
	\subcaptionbox{Packet latency.\label{fig:nebb_plat}}{
		\includegraphics[height=9\baselineskip]{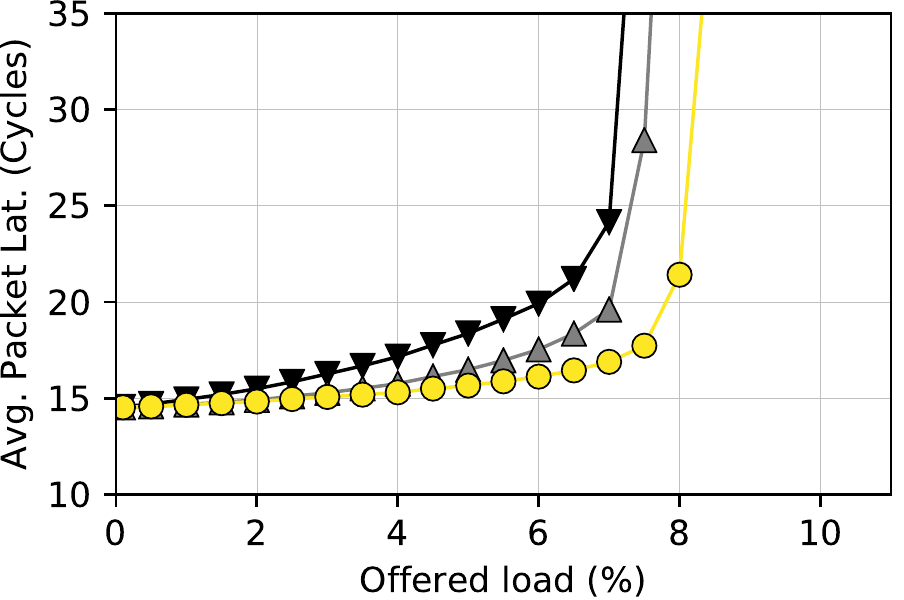}} \hfill
	\subcaptionbox{Buffered flits.\label{fig:nebb_buf_flits}}{
		\includegraphics[height=9\baselineskip]{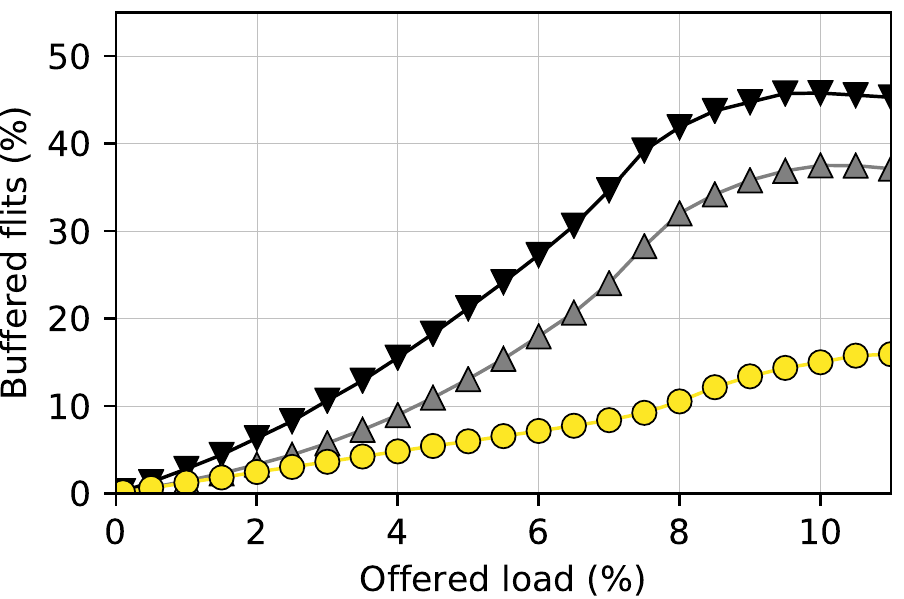}} \hfill
	\subcaptionbox{Router dynamic power.\label{fig:nebb_dyn_power}}{
		\includegraphics[height=9\baselineskip]{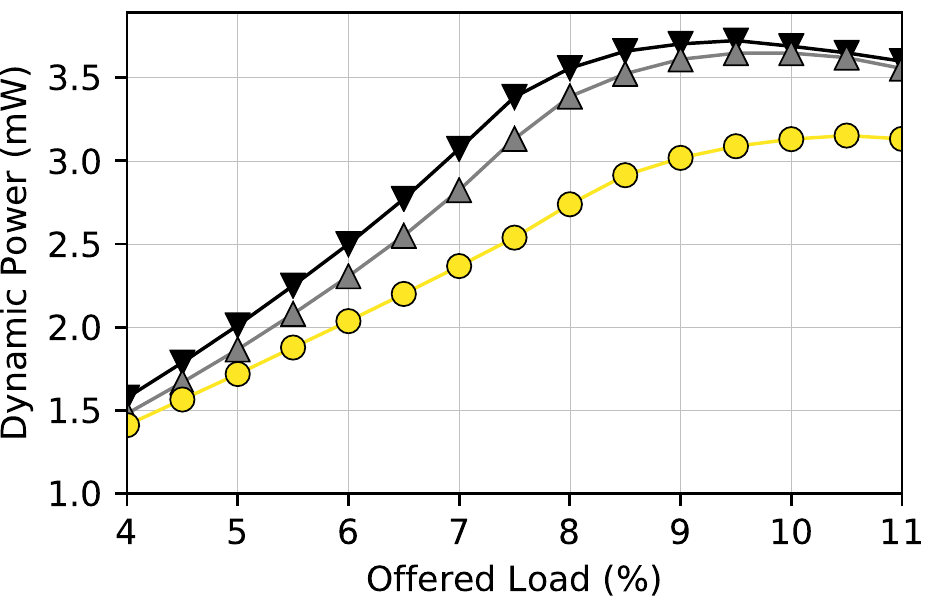}} \hfill
	\caption{$8 \times 8$ $c=4$ mesh performance and efficiency with uniform random single-flit traffic, a DAMQ of 6 flits and 2 VCs.}\label{fig:nebb_single-flit_packets}
\end{figure*}

\begin{table}[t] \caption{Default simulation parameters.}
    \label{table:synth-parameters-bypass} \centering
    \resizebox{\columnwidth}{!}{
    \begin{tabular}{|l|p{0.55\columnwidth}|} \hline
        \multicolumn{2}{|c|}{\textbf{General parameters}} \\ \hline
        \textbf{Topology} & $8 \times 8$ mesh or torus, concentration $c=4$\\ \hline
        \textbf{Link latency} & 1 cycle \\ \hline
        \textbf{Router architecture} & 2/4-stage bypass router \\ \hline
        \textbf{Router size} & 8 ports (4 transit, 4 injection/ejection) \\ \hline
        \textbf{Packet size} & 1 and 5 flits\\ \hline
        \textbf{Buffer implementation} & Shared (DAMQ, \cite{Tamir1992})\\ \hline
        \textbf{Buffer size} & 12 flits (1 private flit per VC)\\ \hline
        \textbf{Routing} & DOR\\ \hline
        \textbf{SA input arbiters} & 8 Round Robin arbiters, \#VCs:1 \\ \hline
        \textbf{SA output arbiters} & 8 Matrix arbiters, 8:1 \\ \hline
        \textbf{LA arbiters} & 8 Matrix arbiters, 8:1\\ \hline
        \textbf{VA policy} & Highest number of credits \\ \hline
        \textbf{Channel width} & 128 bits \\ \hline\hline
       
        \multicolumn{2}{|c|}{\textbf{Synthetic traffic parameters}} \\ \hline
        \textbf{Num. VCs} & 2\\ \hline
        \textbf{Simulation cycles} & 50.000 cycles \\ \hline\hline
        
        \multicolumn{2}{|c|}{\textbf{Full-system parameters}} \\ \hline
        \textbf{CPUs} & 64x O3 ARM @ 2 GHz (DerivO3CPU) \\ \hline
        \textbf{L1 caches} & 32KB (L1-I) and 64KB (L1-D) per core \\ \hline
        \textbf{L2 caches} & 64x 256KB shared banks\\ \hline
        \textbf{Memory controllers} & 16 (first and last mesh rows)\\ \hline
        \textbf{Num. VCs} & 3 (1 per VN) \\ \hline
        \textbf{NoC frequency} & 2 GHz \\ \hline
        \textbf{Simulation cycles} & $10^8$ cycles\\ \hline\hline
       
        \multicolumn{2}{|c|}{\textbf{DSENT parameters}} \\ \hline
        \textbf{Frequency} & 2 GHz \\ \hline
        \textbf{Technology} & Tri-Gate 11nm LVT process \\ \hline
        
    \end{tabular}
    }
\end{table}

We have implemented the router architecture described in
Section~\ref{sect:state_of_the_art} and the bypass schemes described in
Section~\ref{sect:hybrid_flow_control} in BookSim~\cite{Jiang2013}.
We model a 256-core network, arranged as an $8 \times 8$ mesh or torus topology,
with concentration $c=4$. The router employs a two-stage switch allocator similar
to~\cite{kumar2008,krishna2010} to balance pipeline stages.
Priority is given to LAs over buffered flits.
In the non-bypass pipeline, priority is given to body flits as presented in
Section~\ref{sect:switch_allocator}.
Simulation parameters are shown in Table~\ref{table:synth-parameters-bypass}, unless otherwise noted in the text.


We employ six bypass models. \emph{Empty VC+Arb} implements \emph{Empty VC} using an arbiter between LAs. \emph{Baseline} is a WH reference that uses LA conflict-check, while
\emph{Baseline+Arb} is a WH reference using an arbiter between LAs. Additionally, we
implement the three \emph{NEBB} variants introduced in Section~\ref{sect:hybrid_flow_control}:
\emph{NEBB-WH}, \emph{NEBB-VCT} (which also employs VCT in the non-bypass pipeline) and
\emph{NEBB-Hybrid}.

Experiments use synthetic traffic, with single-flit packets or bimodal traffic.
Bimodal traffic resembles a coherence protocol
using packets of one (control) and five (data) flits. A single-flit packet
ratio of 80\% is used~\cite{ma2015,Ma2012}. The traffic pattern employed is
random uniform, but we also evaluate bit-reversal, transpose and hotspot (with hotspots in nodes 0, 15, 240 and 255).
We measure relevant metrics such as average packet latency, dynamic power, and percentage of buffered flits.
The latter divides the amount of times a flit is buffered by the total number of times a flits is forwarded, averaged for all flits.

We use gem5~\cite{binkert2011gem5} to evaluate \emph{NEBB-Hybrid} with real traffic provided by Full-System (FS) simulations. We simulate a tiled system with 64 Out-Of-Order (O3) ARM CPUs with private L1s and a shared L2 distributed among the tiles. We use Virtual Networks (VN) to avoid protocol-deadlock. Each VN has one VC, except for the case of for \emph{Empty VC+Arb} in the torus, which requires 2 VCs to avoid the inherent routing-deadlock of rings when using Dateline~\cite{Dally2003}. We run the PARSEC suite~\cite{bienia2008parsec} with the simlarge input set for every benchmark to have enough workload for the 64 cores. Because of the excessive simulation time of the whole benchmarks, we present results from the first 100 million cycles.

Dynamic power results are obtained using DSENT~\cite{Sun2012}. We have
implemented a model of the bypass router based on the default
four-stage router model of DSENT.
The dynamic power of the buffers and allocators from DSENT is multiplied
by the ratio of buffered flits over all the received flits per router.
The LA arbiters employed are equal to the arbiters in the second stage of the
switch allocator. Therefore, the LA arbiters power equals the power of the second stage of the
switch allocator provided by DSENT. We use no correction factor because these arbiters
are used for every LA, and one LA is received for each flit.
We estimate that the extra control logic (such
as the checks of the packet size, the occupancy of the buffer to bypass, etc)
is negligible compared to the consumption of the buffers, crossbar or
arbiters.

%



\section{Results} \label{Evaluation} \label{sect:evaluation}
\begin{figure*}[t]
	\centering
	\begin{subfigure}[b]{0.75\textwidth}
		\includegraphics[width=\textwidth]{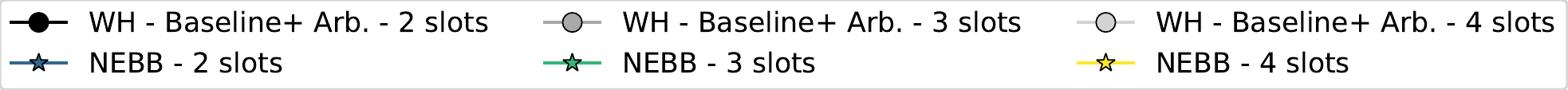}
	\end{subfigure}
	
	\subcaptionbox{Average packet latency.\label{fig:minimal_buffering_packet_latency}}{
		\includegraphics[height=9\baselineskip]{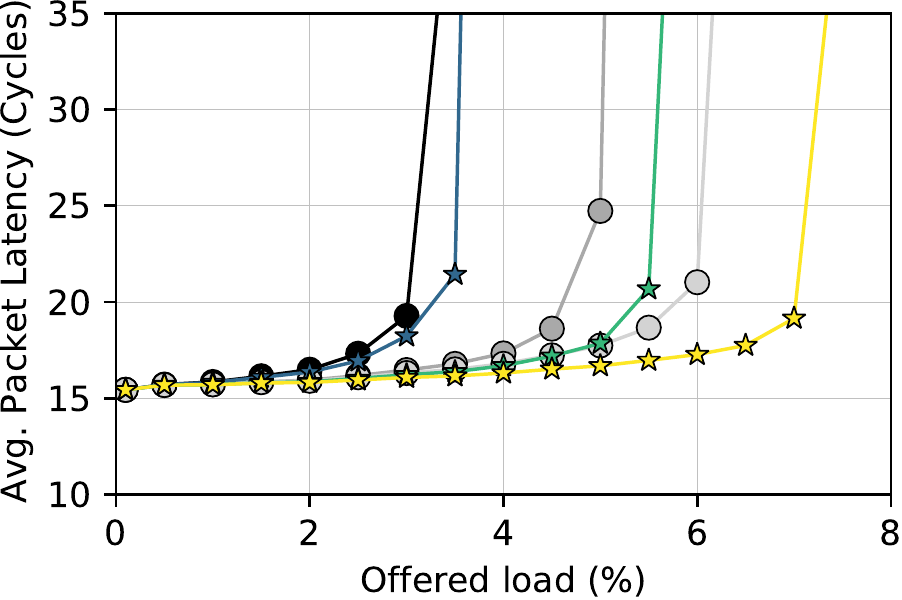}} \hfill
	\subcaptionbox{Throughput.\label{fig:minimal_buffering_throughput}}{
		\includegraphics[height=9\baselineskip]{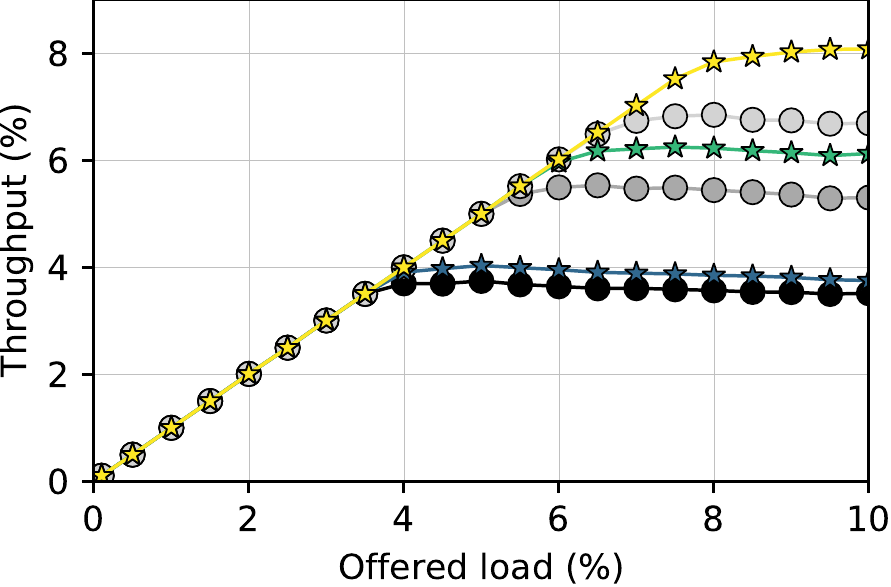}} \hfill
	\subcaptionbox{Buffered flits.\label{fig:minimal_buffering_buf_flits}}{
		\includegraphics[height=9\baselineskip]{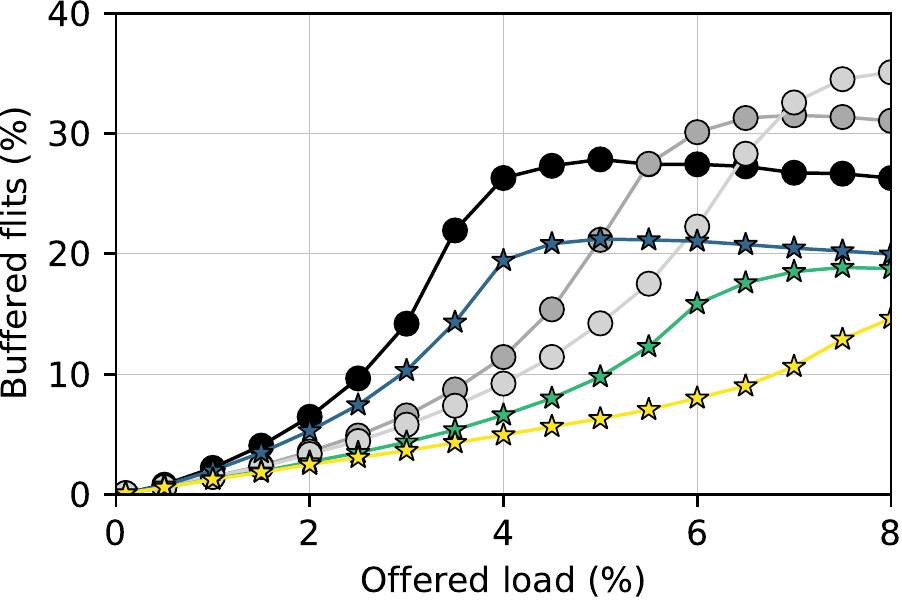}} \hfill
	
	\caption{Performance of bypass routers in an $8 \times 8$ $c=4$ mesh with single-flit random-uniform traffic, an minimal buffering  without VCs.}\label{fig:nebb_minimal_buffering}
\end{figure*}

\begin{figure*}[t]
	\centering
	\begin{subfigure}[b]{0.66\textwidth}
		\includegraphics[width=\textwidth]{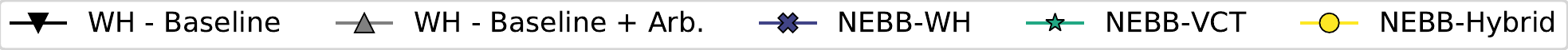}
	\end{subfigure}
	
	\subcaptionbox{Packet latency.\label{fig:nebb_plat-multi}}{
		\includegraphics[height=9\baselineskip]{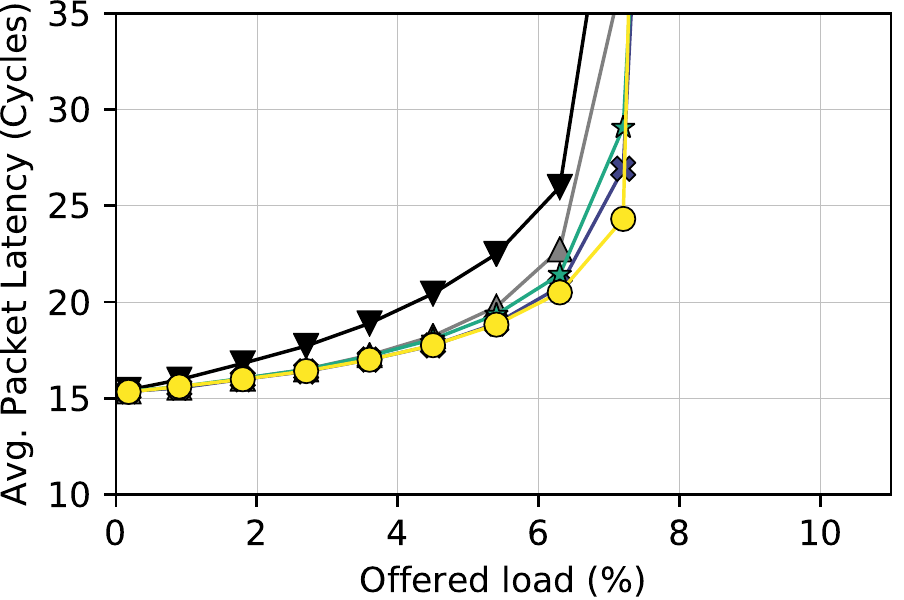}} \hfill
	\subcaptionbox{Buffered flits.\label{fig:nebb_buf_flits-multi}}{
		\includegraphics[height=9\baselineskip]{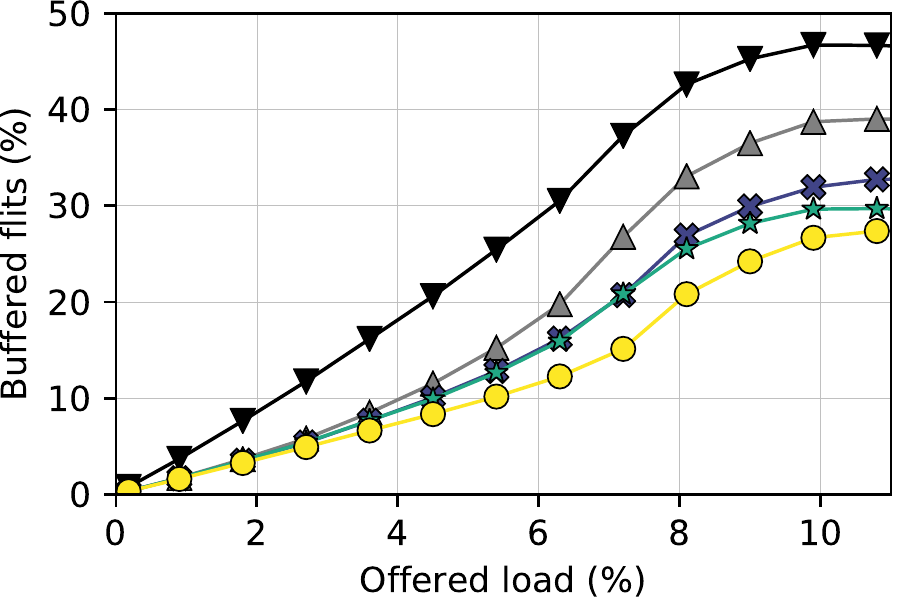}} \hfill
	\subcaptionbox{Router dynamic power.\label{fig:nebb_dyn_power-multi}}{
		\includegraphics[height=9\baselineskip]{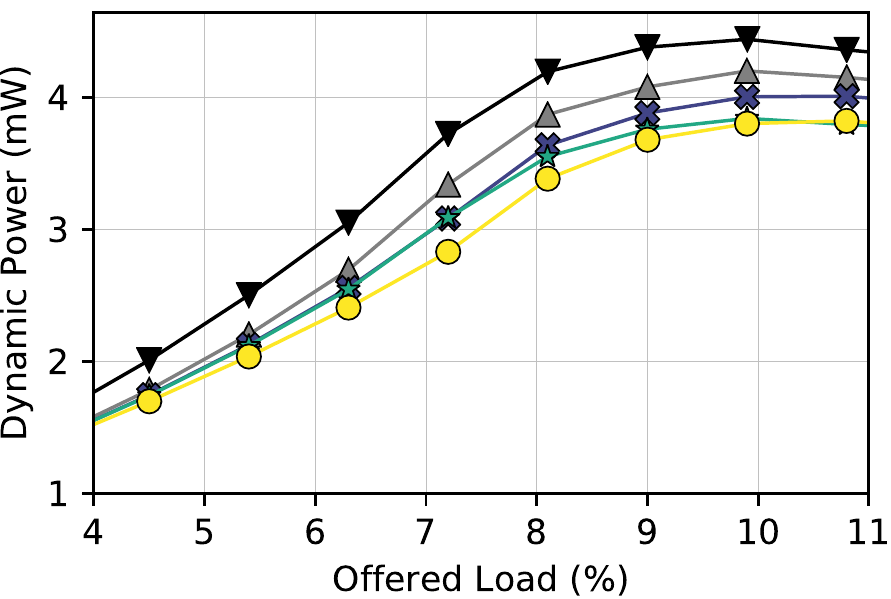}} \hfill
	\caption{$8 \times 8$ $c=4$ mesh performance and efficiency with uniform random and bimodal traffic, a DAMQ of 12 flits and 2 VCs.}\label{fig:bimodal_traffic}
\end{figure*}

\begin{figure*}[t]
    \centering
	\begin{subfigure}[b]{0.66\textwidth}
		\includegraphics[width=\textwidth]{images/f10_legend.pdf}
	\end{subfigure}
	
	\subcaptionbox{Bit-reversal.\label{fig:bitrev}}{
		\includegraphics[height=9\baselineskip]{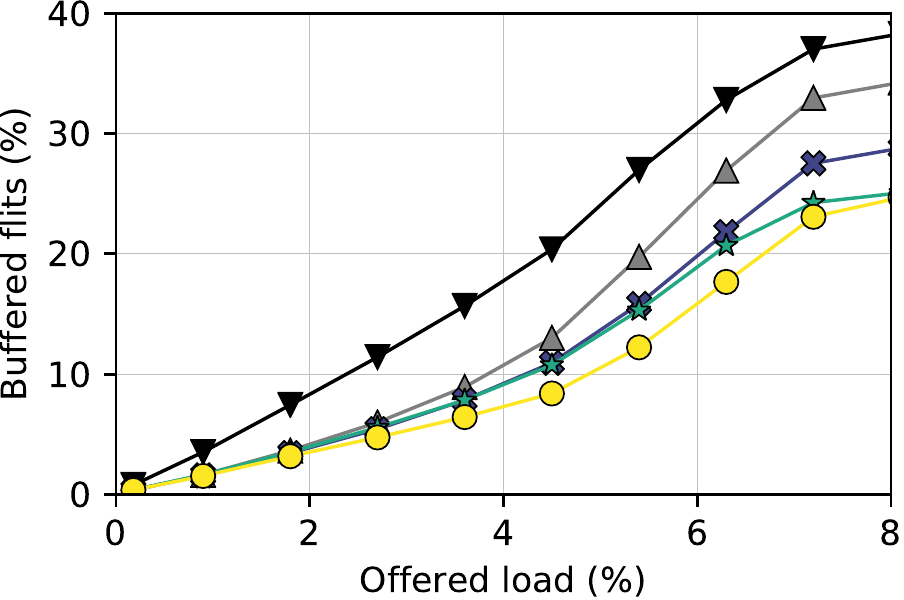}} \hfill
	\subcaptionbox{Transpose.\label{fig:transpose}}{
		\includegraphics[height=9\baselineskip]{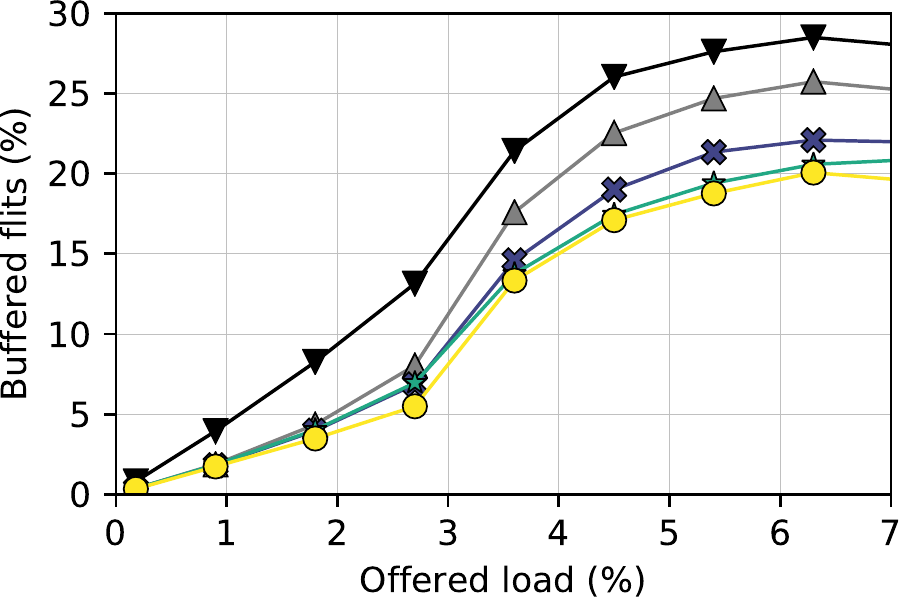}} \hfill
	\subcaptionbox{Hotspot.\label{fig:hotspot}}{
		\includegraphics[height=9\baselineskip]{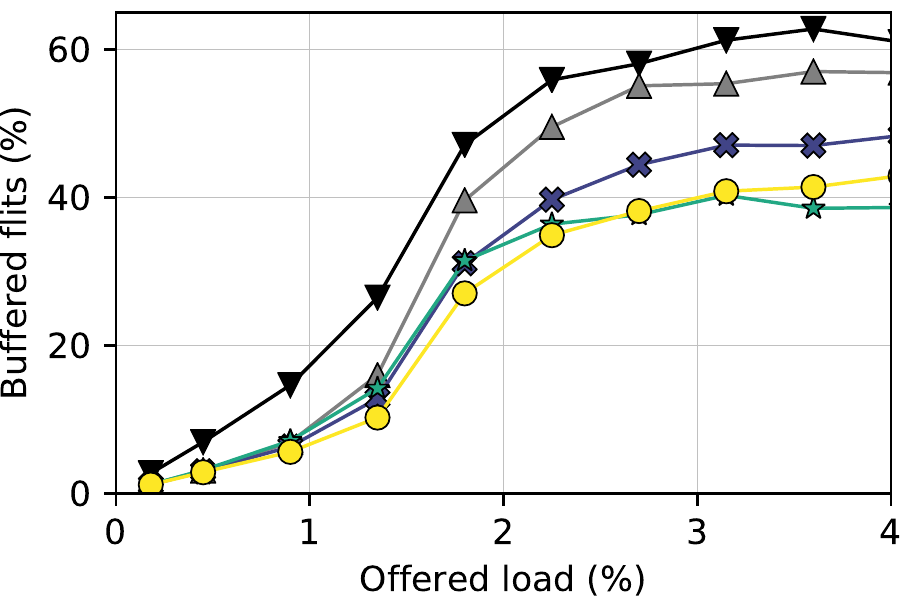}} \hfill
	
	\caption{Buffered flits in an $8 \times 8$ $c=4$ mesh for different traffic patterns, using bimodal traffic, a DAMQ of 12 flits and 2 VCs.}\label{fig:traffic_patterns}
\end{figure*}

\begin{figure*}[t]
	\centering
	\begin{subfigure}[b]{0.66\textwidth}
		\includegraphics[width=\textwidth]{images/f10_legend.pdf}
	\end{subfigure}
	
	\subcaptionbox{Average packet latency.\label{fig:torus_packet_latency}}{
		\includegraphics[height=9\baselineskip]{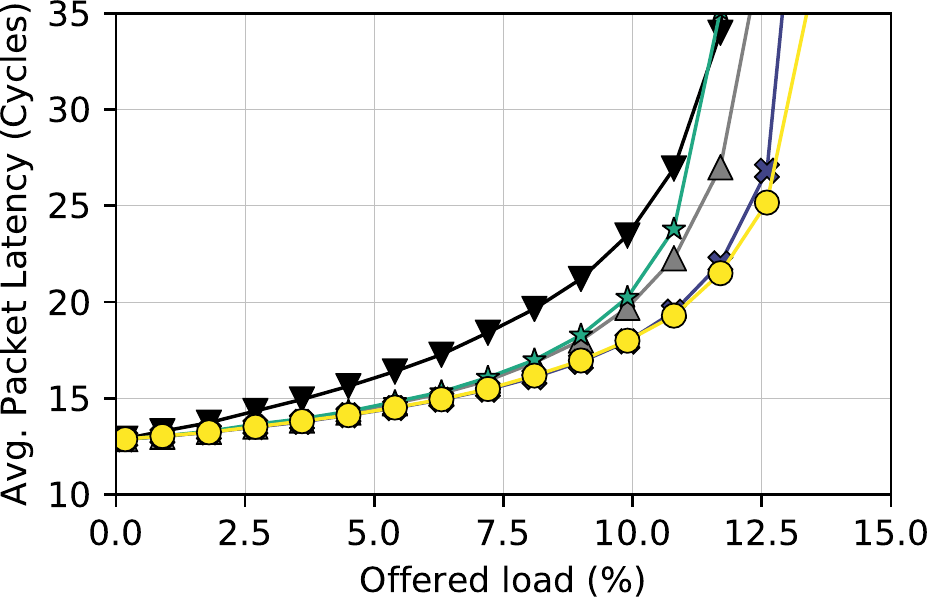}} \hfill
	\subcaptionbox{Buffered flits.\label{fig:torus_buf_flits}}{
		\includegraphics[height=9\baselineskip]{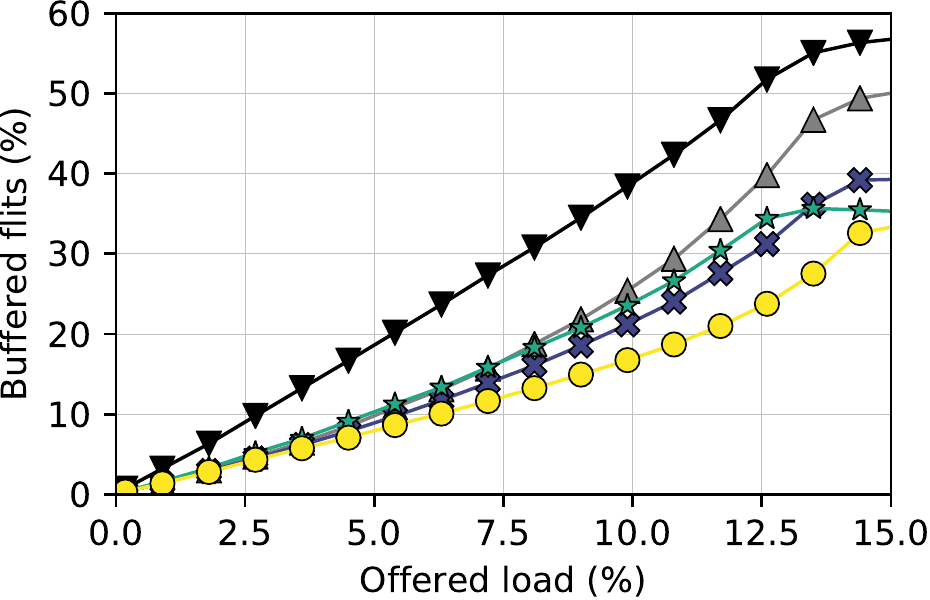}} \hfill
	\subcaptionbox{Router dynamic power.\label{fig:torus_dyn_power}}{
		\includegraphics[height=9\baselineskip]{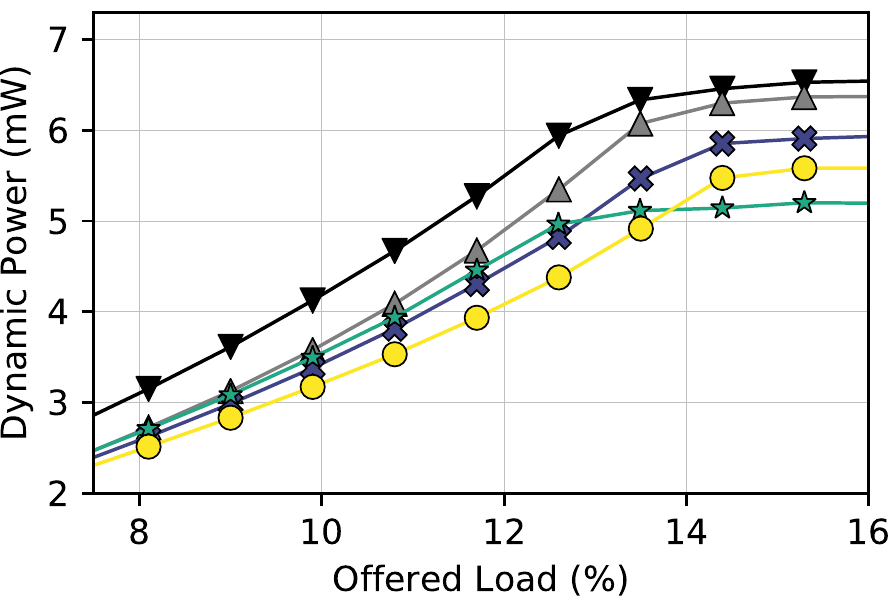}} \hfill
	
	\caption{Performance of an $8 \times 8$ $c=4$ torus with bimodal uniform traffic, a DAMQ of 12 flits and 2 VCs.}\label{fig:torus}
\end{figure*}

\begin{figure*}[t]
	\centering
	\begin{subfigure}[b]{0.66\textwidth}
		\includegraphics[width=\textwidth]{images/f10_legend.pdf}
	\end{subfigure}
	
	\subcaptionbox{Bit-reversal.\label{fig:torus_bitrev}}{
		\includegraphics[height=9\baselineskip]{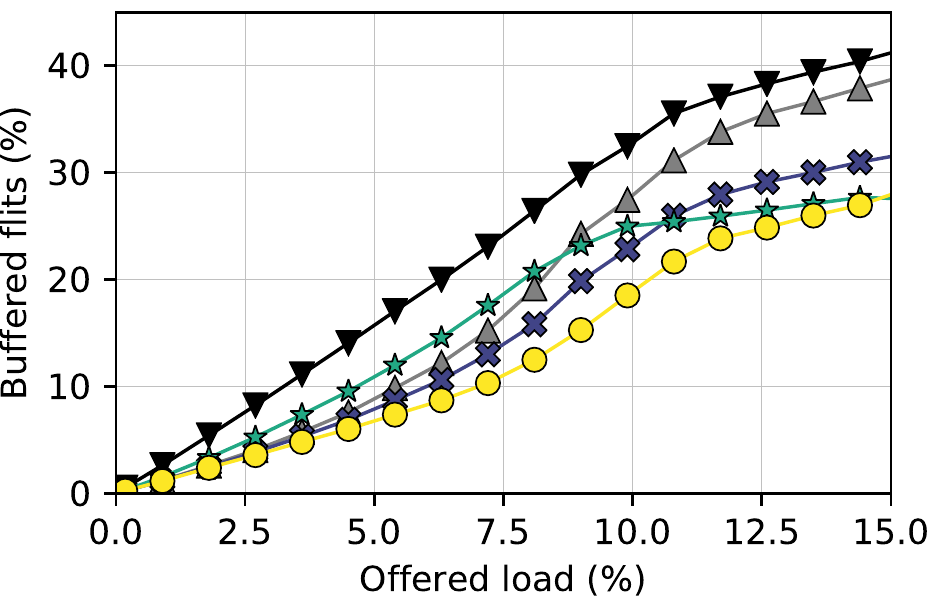}} \hfill
	\subcaptionbox{Transpose.\label{fig:torus_transpose}}{
		\includegraphics[height=9\baselineskip]{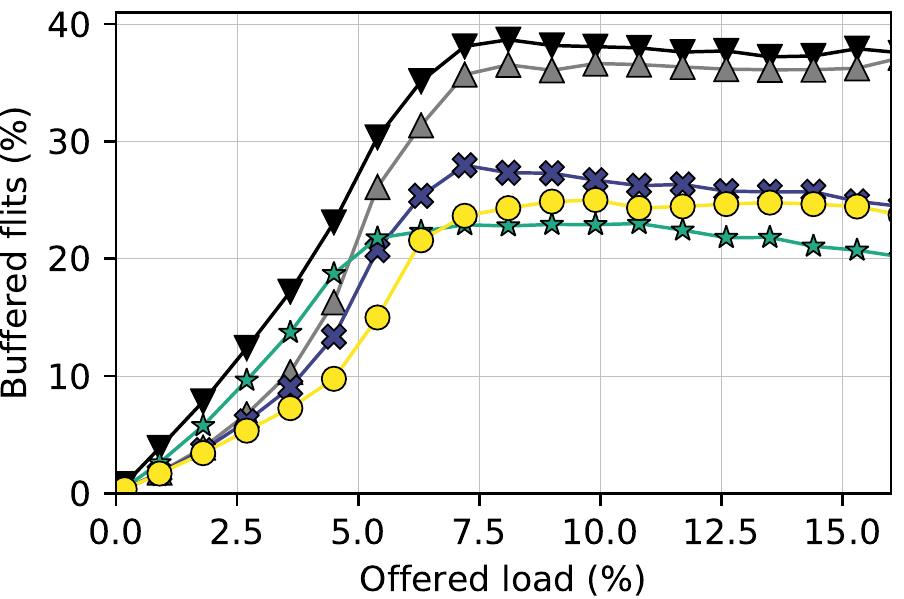}} \hfill
	\subcaptionbox{Hotspot.\label{fig:torus_hotspot}}{
		\includegraphics[height=9\baselineskip]{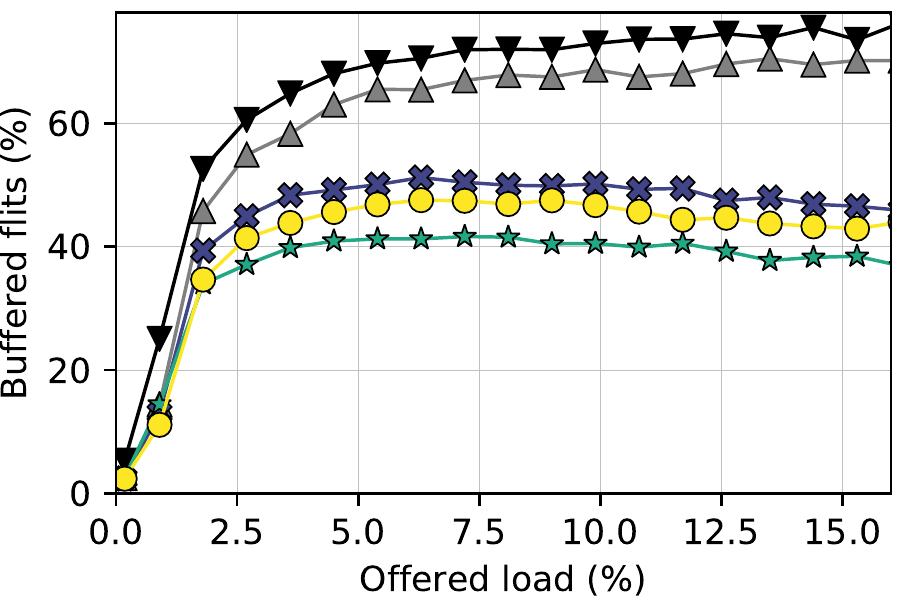}} \hfill
	
	\caption{Buffered flits in an $8 \times 8$ $c=4$ torus for different traffic patterns, using bimodal traffic, a DAMQ of 12 flits and 2 VCs.}\label{fig:torus_traffic_patterns}
\end{figure*}


This section highlights the most notable empirical results when comparing the different router architectures. A performance comparison between the \emph{Empty VC} mechanism and our WH baseline. Next, the performance and efficiency of \emph{NEBB} using single-flit packets is considered; Bimodal traffic in mesh and torus is also evaluated. Finally, a sensitivity analysis is carried out.

\subsection{Empty VC vs WH Flow Control}\label{sect:eval_on/off_wh}


Figure~\ref{fig:on_off} compares packet latency of two implementation of WH flow control, with the restriction of an empty buffer to forward data (\emph{Empty VC}) or without it (\emph{WH-Baseline}).  Both versions have LA arbiters; this is the configuration with the best performance, as observed in the results presented in Section~\ref{sect:nebb_single-flit_packets}. Bimodal traffic is used in all cases.

Figures~\ref{fig:on_off_shared_buf_size_10} and~\ref{fig:on_off_shared_buf_size_20} employ shared buffers (DAMQ) with space for 10 and 20 flits, respectively. Clearly, the WH version has better results in respect of \emph{Empty VC} with 1, 2 and 4 VCs. It is particularly notorious the low throughput of \emph{Empty VC} with 1 and 2 VCs. \emph{WH-Baseline} with a single VC is even better than \emph{Empty VC} with 4 VCs. With 8 or more VCs, the improvement of \emph{WH-Baseline} runs out, providing similar results in both configurations. 
Note that there are no results with 16 VCs in Figure~\ref{fig:on_off_shared_buf_size_10} because each VC needs a private slot and the DAMQ only has 10 flits.

Figures~\ref{fig:on_off_private_buf_size_5} and~\ref{fig:on_off_private_buf_size_10} have a private buffer per VC with room for 5 and 10 flits, respectively. Again, \emph{WH-Baseline} is better for 1 to 4 VCs, and similar with 8 and 16 VCs. 

Reducing the amount of buffering space is critical to reduce power consumption and area in NoC designing. \emph{WH-Baseline} is clearly better than \emph{Empty VC} with a low number of VCs and a low amount of buffering space. Therefore, we use this configuration as our baseline in the next sections.

\subsection{NEBB using Single-Flit Packets}\label{sect:nebb_single-flit_packets}


Figure~\ref{fig:nebb_single-flit_packets} compares packet latency, buffered flits and dynamic power of each bypass mechanism; in all cases, lower values are better. These first evaluations use single-flit traffic, so all \emph{NEBB} variants are equivalent. Only 6 slots per shared buffer are used, adapted to the small packet size.

The amount of buffered flits in~\ref{fig:nebb_buf_flits} grows with the network load.
\emph{WH-Baseline} buffers flits when the buffers are non-empty or there are LA conflicts (all the conflicting LAs are discarded). The \emph{WH-Baseline+Arb} model is similar, but one LA proceeds in case of conflicts, reducing the use of buffers. In \emph{NEBB}, buffers are only used when conflicts occur, not because of non-empty buffers, minimizing the buffer utilization. This translates into latency and power savings, particularly at intermediate loads.
At 7\% load (0.07 flits/node/cycle), \emph{NEBB} reduces \emph{WH-Baseline's} latency in 30.1\% and buffered flits in 75.9\%. From these values, 18.8\% and 30.7\% respectively come from the LA arbiter, as observed in \emph{WH-Baseline+Arb} results. Regarding dynamic power, \emph{NEBB} saves 23.0\% over \emph{WH-Baseline} at 7\% load.

Figure~\ref{fig:nebb_minimal_buffering} shows the latency, throughput and buffered flits of \emph{WH-Baseline + Arb.} and \emph{NEBB} using routers with minimal buffering. Evaluated configurations do not have VCs (i.e. equivalent to 1 VC) and the input buffers have 2, 3 or 4 slots. Injected packets only have one flit; in this case, we do not plot results using 5-flit packets because NEBB-VCT cannot be applied.
\emph{NEBB} reduces the number of buffered flits under these conditions by 24.5\%, 39.3\% and 51.3\% for 2, 3, and 4 slots respectively. As a consequence of reducing Head of Line Blocking (HoLB) and the pressure over the buffers, \emph{NEBB} achieves 6.8\%, 15.5\%, 20.8\% more throughput than \emph{WH-Baseline + Arb.} for 2, 3 and 4 slots, respectively.

\subsection{\emph{NEBB} Flow Control and \emph{Hybrid}}\label{sect:nebb_bimodal}

Figure~\ref{fig:bimodal_traffic} compares the \emph{NEBB} alternatives using bimodal traffic.
The three \emph{NEBB} variants outperform the baselines, and \emph{Hybrid} presents the best
results since it maximizes the cases in which bypass is used.

Both \emph{NEBB-WH} and \emph{NEBB-VCT} present similar results.
VCT has slightly higher latency and lower throughput, which translate
into slightly lower power results after saturation.

\emph{NEBB-Hybrid} has the best results in latency, buffered flits and dynamic power with
a reduction of 20.6\%, 60.1\% and 21.1\%, respectively, over \emph{WH-Baseline} at a load around 6\%.

Figure~\ref{fig:traffic_patterns} depicts the buffer utilization for different traffic patterns. The results are similar to the previous ones with uniform random traffic, with \emph{NEBB} mechanisms improving the utilization of the bypass and \emph{Hybrid} being the optimal version.

\subsection{\emph{NEBB} in Torus networks}

Figure~\ref{fig:torus} depicts results of the bypass router in a torus. All the bypass
mechanisms use FBFC except for \emph{NEBB-VCT}, which relies on Bubble flow
control~\cite{Carrion1997} since it only employs VCT.

In general, base latency is lower than in the mesh, and throughput is almost doubled,
proving that FBFC and bypass routers with shared buffers operate correctly together.
FBFC presents better results than Bubble, with \emph{NEBB-VCT} presenting worse latency
than \emph{WH-Baseline+Arb}. Once more, \emph{NEBB-Hybrid} has the best results,
improving latency over \emph{WH-Baseline} by 28.4\% and dynamic power by 24.4\% at 11\% load.

Figure~\ref{fig:torus_traffic_patterns} shows the buffer utilization applying different traffic patterns. \emph{NEBB} always writes less flits in the buffers than the baselines. In general, \emph{NEBB-Hybrid} is the best choice between the \emph{NEBB} mechanisms, except in the case of the hotspot pattern, in which \emph{NEBB-VCT} is slightly better.

\begin{figure*}[t]
	\centering
	\begin{subfigure}[b]{\columnwidth}
		\includegraphics[width=\textwidth]{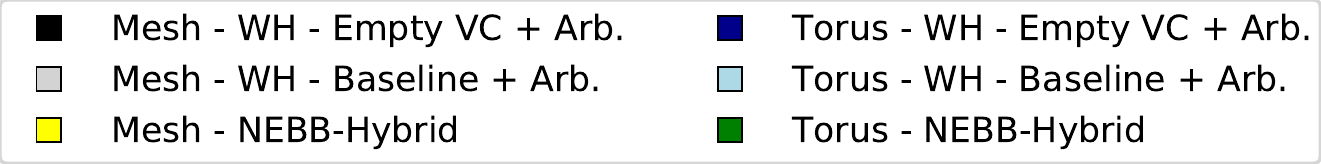}
	\end{subfigure}

	\begin{subfigure}[b]{\columnwidth}
		\includegraphics[width=\columnwidth]{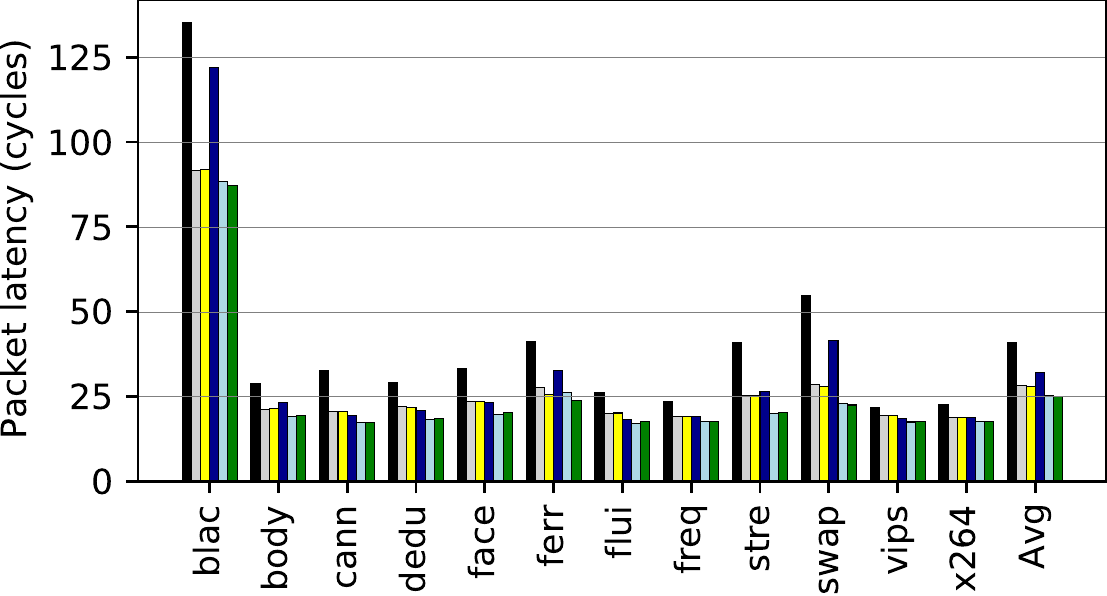}
	    \caption{Avg. packet latency}\label{fig:fs_avg_plat}
	\end{subfigure}
	\begin{subfigure}[b]{\columnwidth}
		\includegraphics[width=\columnwidth]{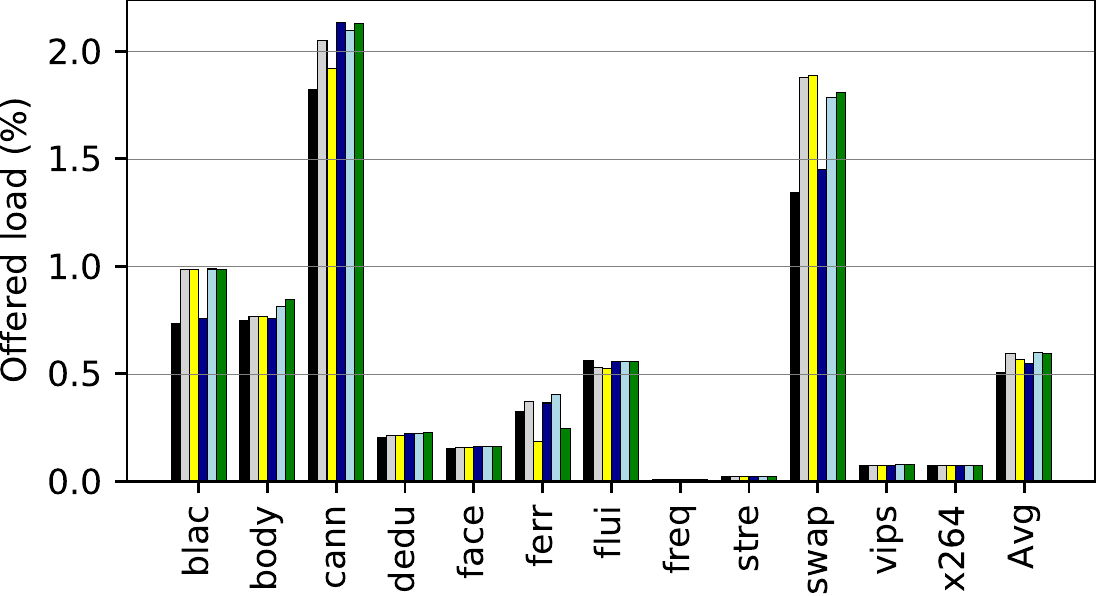}
	    \caption{Offered load}\label{fig:fs_offered_load}
	\end{subfigure}
	
	\caption{Real-traffic performance.}\label{fig:fs_evaluation}
\end{figure*}

\begin{figure*}[t]
	\centering
	\begin{subfigure}[b]{\textwidth}
		\includegraphics[width=\textwidth]{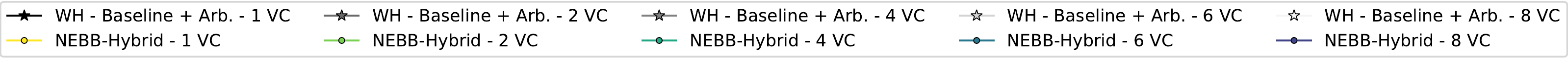}
	\end{subfigure}
	
	\subcaptionbox{DAMQ. Total size: 10 flits.\label{fig:shared_buf_size_10}}{
		\includegraphics[height=9\baselineskip]{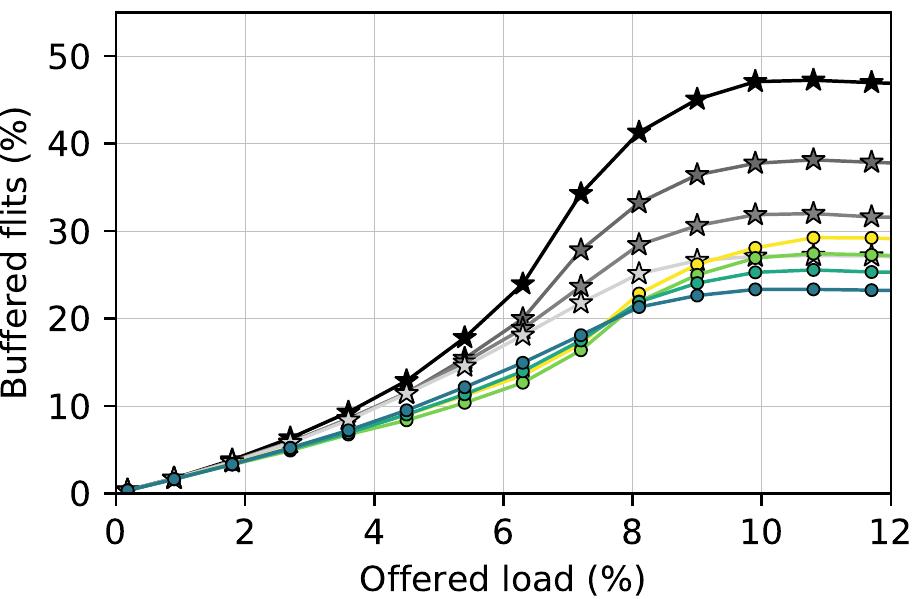}}\hspace{1cm}
	\subcaptionbox{DAMQ. Total size: 20 flits.\label{fig:shared_buf_size_20}}{
		\includegraphics[height=9\baselineskip]{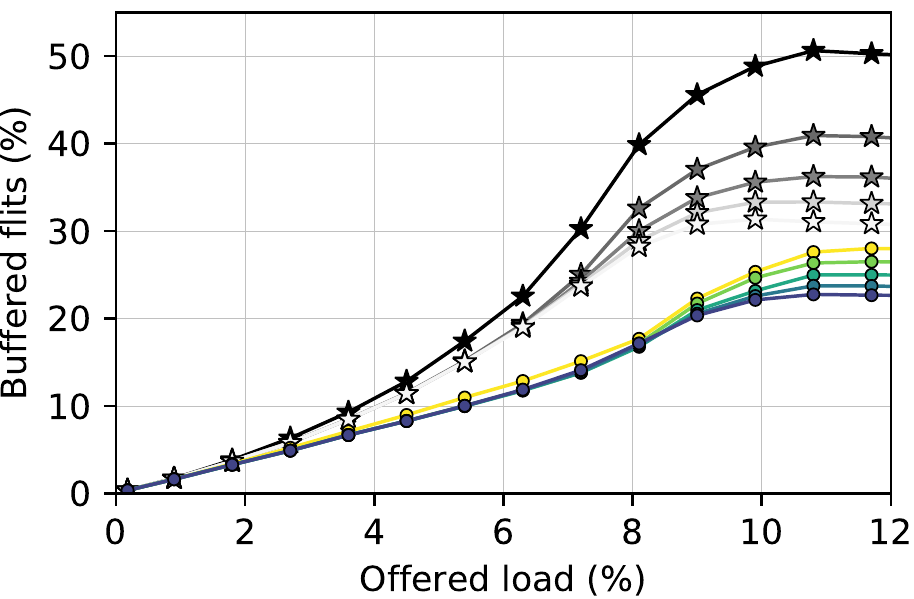}}
	
	\subcaptionbox{Private VC. VC size: 5 flits.\label{fig:private_buf_size_5}}{
		\includegraphics[height=9\baselineskip]{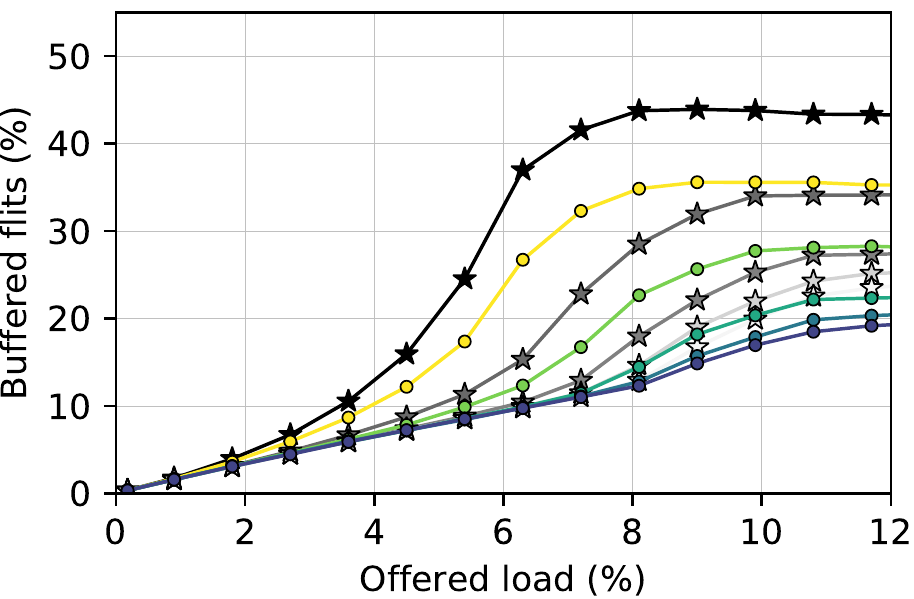}}\hspace{1cm}
	\subcaptionbox{Private VC. VC size: 10 flits.\label{fig:private_buf_size_10}}{
		\includegraphics[height=9\baselineskip]{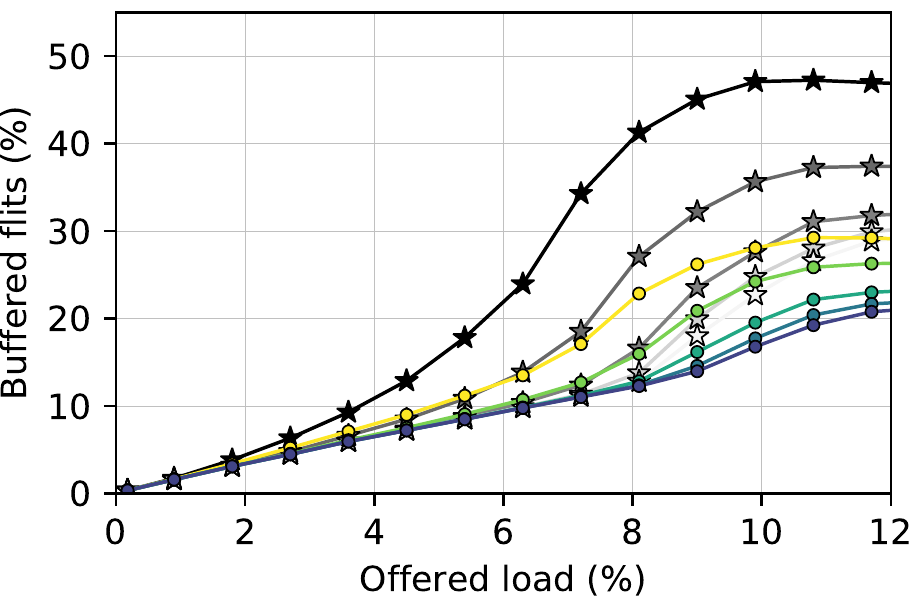}}
	
	\caption{Buffer utilization for a mesh with different number of VCs and buffer sizes using bimodal traffic.}\label{fig:vcs_buf_size}
\end{figure*}

\begin{figure*}[t]
	\centering
	\begin{subfigure}[b]{0.38\textwidth}
		\includegraphics[width=\textwidth]{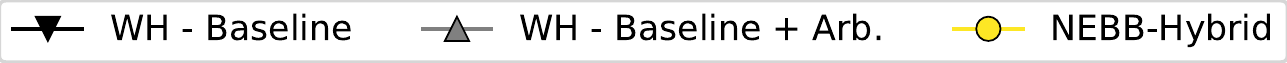}
	\end{subfigure}
	
	\subcaptionbox{Buffered flits. Priority to flits.\label{fig:priority_flits}}{
        \includegraphics[height=8.5\baselineskip]{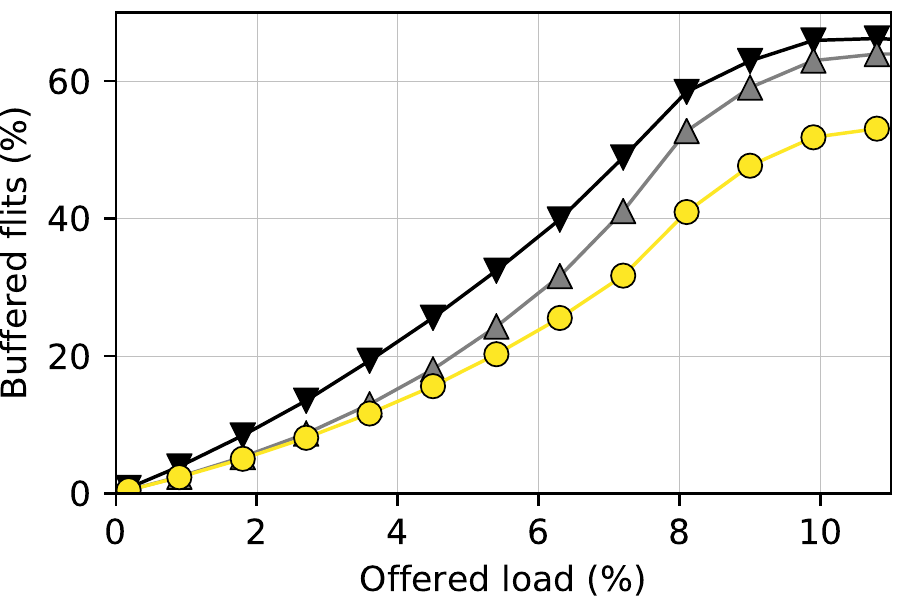}}\hfill
	\subcaptionbox{Throughput. Priority to flits.\label{fig:throughput_priority_flits}}{
        \includegraphics[height=8.5\baselineskip]{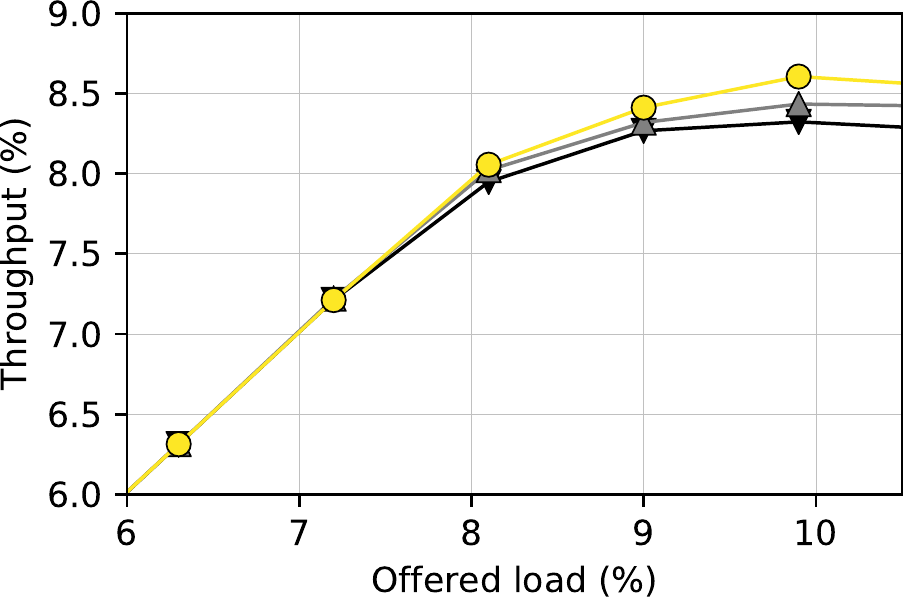}}\hfill
	\subcaptionbox{Latency histogram in saturation (load 10\%). Priority to flits.\label{fig:hist_priority_flits}}{
        \includegraphics[height=8.5\baselineskip]{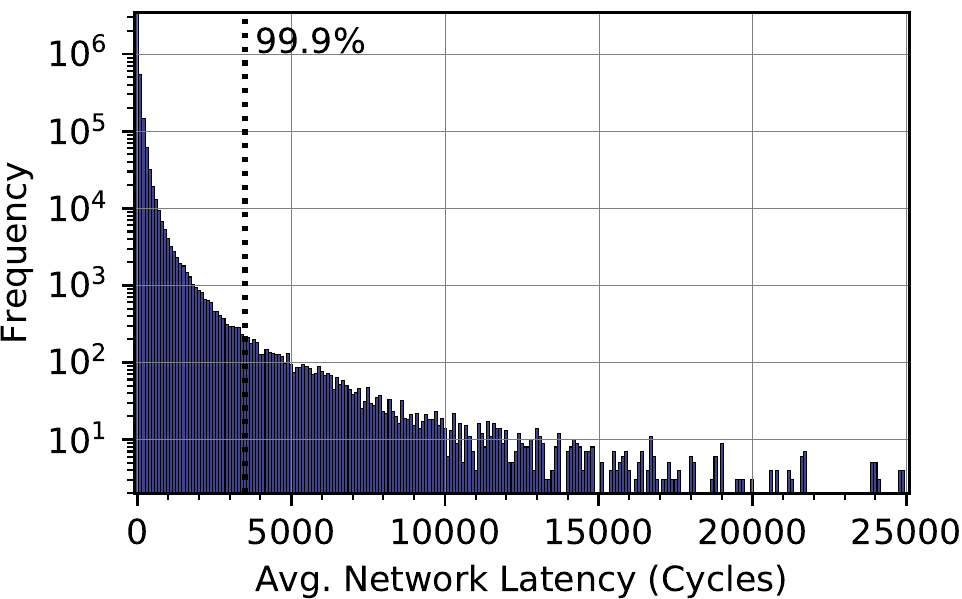}}

	\subcaptionbox{Buffered flits. Priority to LAs.\label{fig:priority_lookaheads}}{
        \includegraphics[height=8.5\baselineskip]{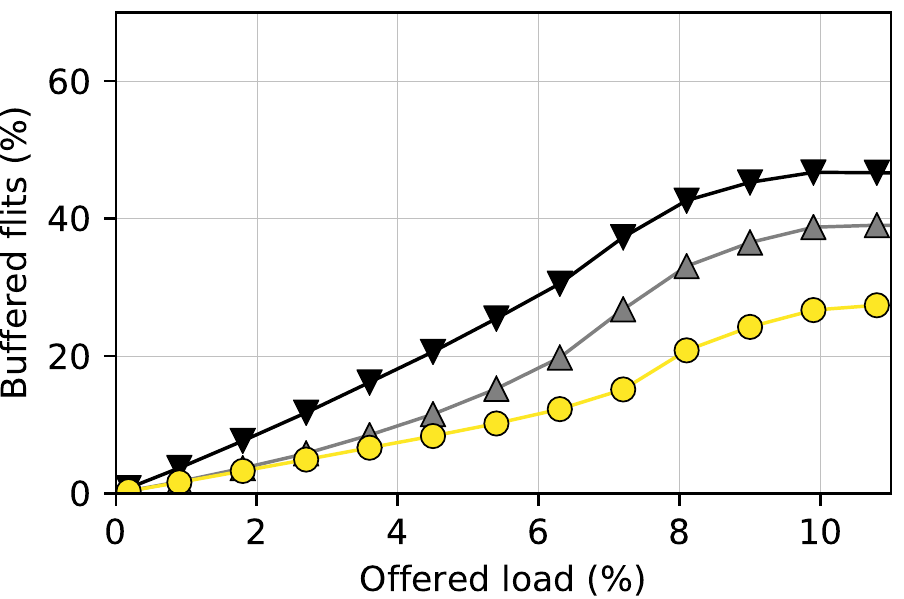}}\hfill
	\subcaptionbox{Throughput. Priority to LAs.\label{fig:throughput_priority_lookaheads}}{
        \includegraphics[height=8.5\baselineskip]{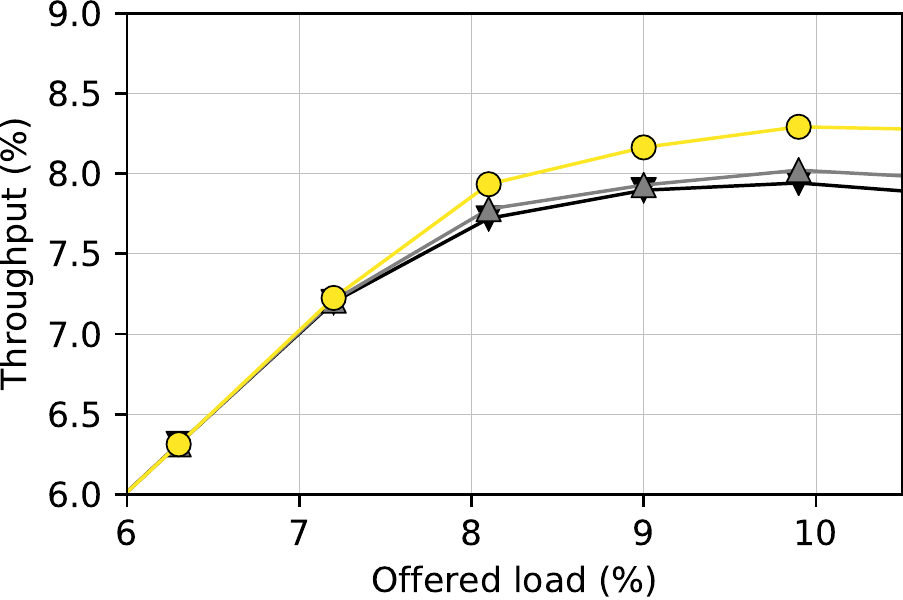}}\hfill
	\subcaptionbox{Latency histogram in saturation (load 10\%). Priority to LAs.\label{fig:hist_priority_lookaheads}}{
        \includegraphics[height=8.5\baselineskip]{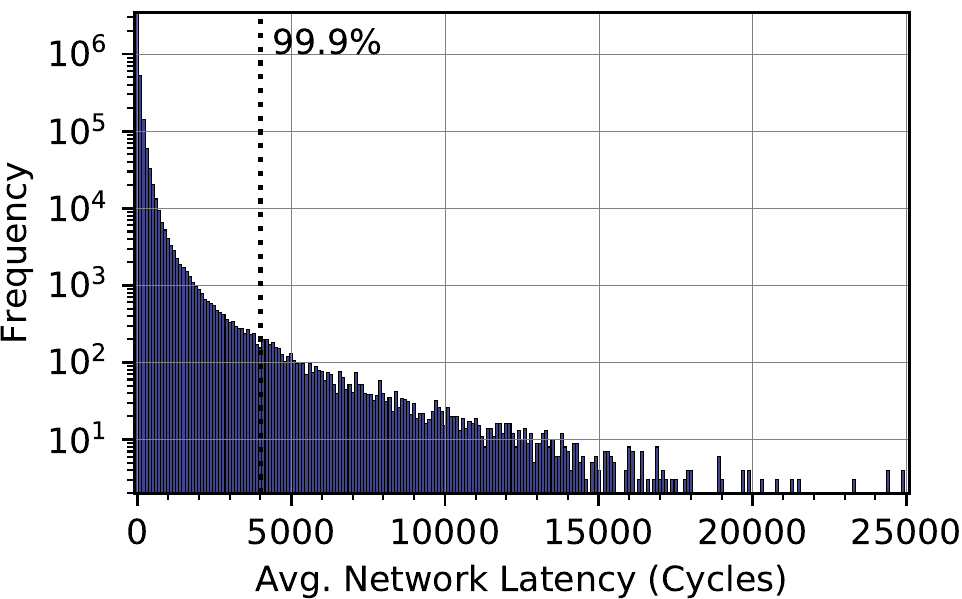}}
	
	\caption{\emph{NEBB-Hybrid} buffer utilization, throughput and network latency histograms prioritizing buffered flits or LAs in case of conflicts.}\label{fig:crossbar_priority}
\end{figure*}

\subsection{Real traffic analysis}\label{sect:real_traffic_analysis}

This section evaluates \emph{NEBB-Hybrid} with real traffic through FS simulations in meshes and tori. We compare \emph{NEBB-Hybrid} with \emph{WH-Empty VC+Arb} and \emph{WH-Baseline+Arb}. All configurations have shared buffers of 12 flits with 3 VCs, 1 VC per VN to avoid protocol-deadlock, except for \emph{Torus-WH-Empty VC} that has 6 VCs of 5 slots each. The reason is that it does not support FBFCL, so it it uses Dateline to avoid routing-deadlock, requiring 2 VCs per VN.

Figure~\ref{fig:fs_evaluation} shows packet latency and offered load. First, the offered load in these benchmarks is very low, around 2\% in \emph{canneal}, the benchmark with the highest load. For this reasons the differences between \emph{WH-Baseline+Arb} and \emph{NEBB-Hybrid} are small as we have shown in the previous sections with synthetic traffic. Second, the configurations that use a torus reduce latency with respect to their mesh counterpart. Third, the \emph{WH-Empty VC+Arb} configurations have on average 44.9\% and 27.3\% more latency in the mesh and torus, respectively, compared with \emph{WH-Baseline+Arb}. This is because \emph{WH-Empty VC+Arb} requires more VCs to offer the same performance as shown in Section~\ref{sect:eval_on/off_wh}.
In the case of the torus, it is remarkable that the FBFCL configurations get lower latency with half the VCs and require less overall buffer space due to Dateline's inefficient VCs utilization.

\subsection{Sensitivity analysis: buffer depth, starvation}\label{sect:nebb_analysis}

\subsubsection{Buffer depth and number of VCs}\label{sect:buffer_depth_and_number_of_VCs}

Figure~\ref{fig:vcs_buf_size} depicts the buffer utilization for \emph{WH-Baseline+Arb} and \emph{Hybrid} with different combinations of VCs and buffer sizes.
Each curve represents the same configuration with a different number of VCs, either sharing the same buffer space (\ref{fig:shared_buf_size_10} and~\ref{fig:shared_buf_size_20}) or using private buffers per VC (\ref{fig:private_buf_size_5} and~\ref{fig:private_buf_size_10}).
With shared buffers, \emph{Hybrid} clearly outperforms \emph{WH-Baseline+Arb}, particularly when the shared buffer size is not very small. With 20 flits per port, no amount of VCs in \emph{WH-Baseline+Arb} matches the result of \emph{Hybrid}. The amount of VCs used in \emph{Hybrid} has a small impact on buffered flits.

In the private buffers evaluations in~\ref{fig:private_buf_size_5} and~\ref{fig:private_buf_size_10} the total amount of storage increases with the VC count.
If buffers are very small (\ref{fig:private_buf_size_5}, buffer per VC equals the maximum packet size of 5 flits) \emph{Hybrid} is better than \emph{WH-Baseline+Arb} for the same number of VCs, but the improvement is modest. Indeed, this is the minimum buffer size for \emph{Hybrid} to use VCT. With larger buffers in~\ref{fig:private_buf_size_10}, the results of \emph{Hybrid} with half the VCs approximately match the result of \emph{WH-Baseline+Arb} before saturation, and get better after this point.

\subsubsection{Crossbar priority to buffered or bypassed flits}\label{sect:crossbar_priority_to_buffered_or_bypassed_flits}

The original bypass condition 2 in Section~\ref{sect:bypass_intro} prioritizes flits in the non-bypass pipeline, but the opposite priority (to LAs) is used in this paper. Figure~\ref{fig:crossbar_priority} compares both alternatives, depicting buffered flits and packet latency histograms. On the one hand, giving priority to LAs decreases significantly the number of buffered flits in all mechanisms, particularly at medium and high load. On the other hand, the maximum throughput decreases slightly.

The packet latency histograms of \emph{Hybrid} show that the number of high latency packets increases with priority to LAs. This issue is shared by all mechanisms when more than 1 VC is used. To reduce peak latency, priority may be given to buffered flits after a given number of cycles. The specific threshold used (e.g. 30 cycles in ~\cite{kumar2008}) presents a tradeoff between the results with priority to LAs or to buffered flits.

\section{Related Work} \label{sect:related}
Sections~\ref{sect:introduction} and~\ref{sect:state_of_the_art} have already presented LookAhead~\cite{Galles1997} and bypass~\cite{kumar2007} mechanisms.
Token Flow Control (TFC,~\cite{kumar2008}) communicates information about the availability of resources among nodes
in a neighborhood. The objective of the mechanism is the improvement of the bypass utilization by choosing low congested
paths, exploiting path diversity with adaptive routing.

Other works attack the ordering issues in NOCs. Scorpio~\cite{daya2014scorpio} implements a globally ordered mesh for snoopy coherence protocols that addresses the storage scalability problem of directory-based protocols. In general terms, it solves the global message order required by snoopy protocols (which is not intrinsic to mesh topologies) with a broadcast network, which is used to inform the Network Interface Controllers (NICs) about the correct order of the messages. When the protocol messages are received in the NICs, they are reorganized according to the previous information.
Since the micro-architecture of the routers used in Scorpio is similar to the one describe in Section~\ref{sect:state_of_the_art}, we presume that it can be combined with NEBB to enhance performance.


Our \emph{Hybrid} approach combines two flow controls, WH and VCT. 
Whole Packet Forwarding (WPF,~\cite{Ma2012}) applies packet-based flow control in a WH network, 
but they do it to relax VC re-allocation requirements in deadlock-free fully adaptive routing NoCs, 
without considering bypass.
In~\cite{Jafri2010} they suggest using two different types of flow control, 
buffered and unbuffered, but again without considering bypass.

The mechanism in SMART~\cite{krishna2013breaking}, implemented in the OpenSMART NoC generator~\cite{Kwon2017},
extends bypass mechanisms to skip multiple routers in a single cycle, multiplying the savings
in latency and power. The bypass conditions need to be satisfied in all the routers in the path.
For this reason, we believe that our \emph{Hybrid} bypass mechanism can be applied to such designs.
A detailed analysis of multi-hop bypass mechanisms based on \emph{NEBB} is left for future work. Preliminary results have been reported in~\cite{perez2019smart++} with promising improvements in terms of the number of VCs and the required buffer size.

ShortPath~\cite{Psarras2016} proposes an alternative pipeline organization for bypass routers that enhances performance by avoiding the speculative allocation of this kind of routers. Its architecture minimizes the time spent by each flit in a router by allowing the bypass of part of the allocation pipeline stages, i.e. only the first stage even when the LA loses in the second stage. As far as we know, ShortPath uses WH (i.e. flit level flow control) as it performs SA for body flits, and allows the storage of multiple packets in the same input VC buffer. 
Combining ShortPath with NEBB is an interesting idea for future work that may increase the utilization of the bypass paths.

\section{Conclusions} \label{sect:conclusions}
Bypass reduces packet latency and power consumption,
which are key aspects of NoC designs.
Our Non-Empty Buffer Bypass proposal is based on a proper analysis
and relaxing of the original bypass conditions.
Variants of \emph{NEBB} following WH and VCT rules are introduced,
and the combination of them, denoted \emph{Hybrid},
maximizes the utilization of the bypass.

We show the effectiveness of \emph{NEBB} and \emph{Hybrid} in
a mesh and a torus using FBFC.
Our proposals decrease by up to 30\% packet latency and up
to 23\% dynamic power savings in relation to applying WH flow control. Additionally, results show that
\emph{Hybrid} outperforms prior proposals with shared buffers,
and requires half the VCs for the same result with private buffers,
what simplifies VC allocation.

Altogether, these results present  \emph{Hybrid} as a competitive
and cost-effective alternative to improve the design and performance
of the NoC bypass techniques in NoC routers.

\bibliographystyle{ieeetr}
\bibliography{ref}

\clearpage
\section*{Biography} \label{sect:biography}

\vskip -2\baselineskip plus -1fil

\begin{IEEEbiography}[{\includegraphics[width=1in,height=1.25in,clip,keepaspectratio]{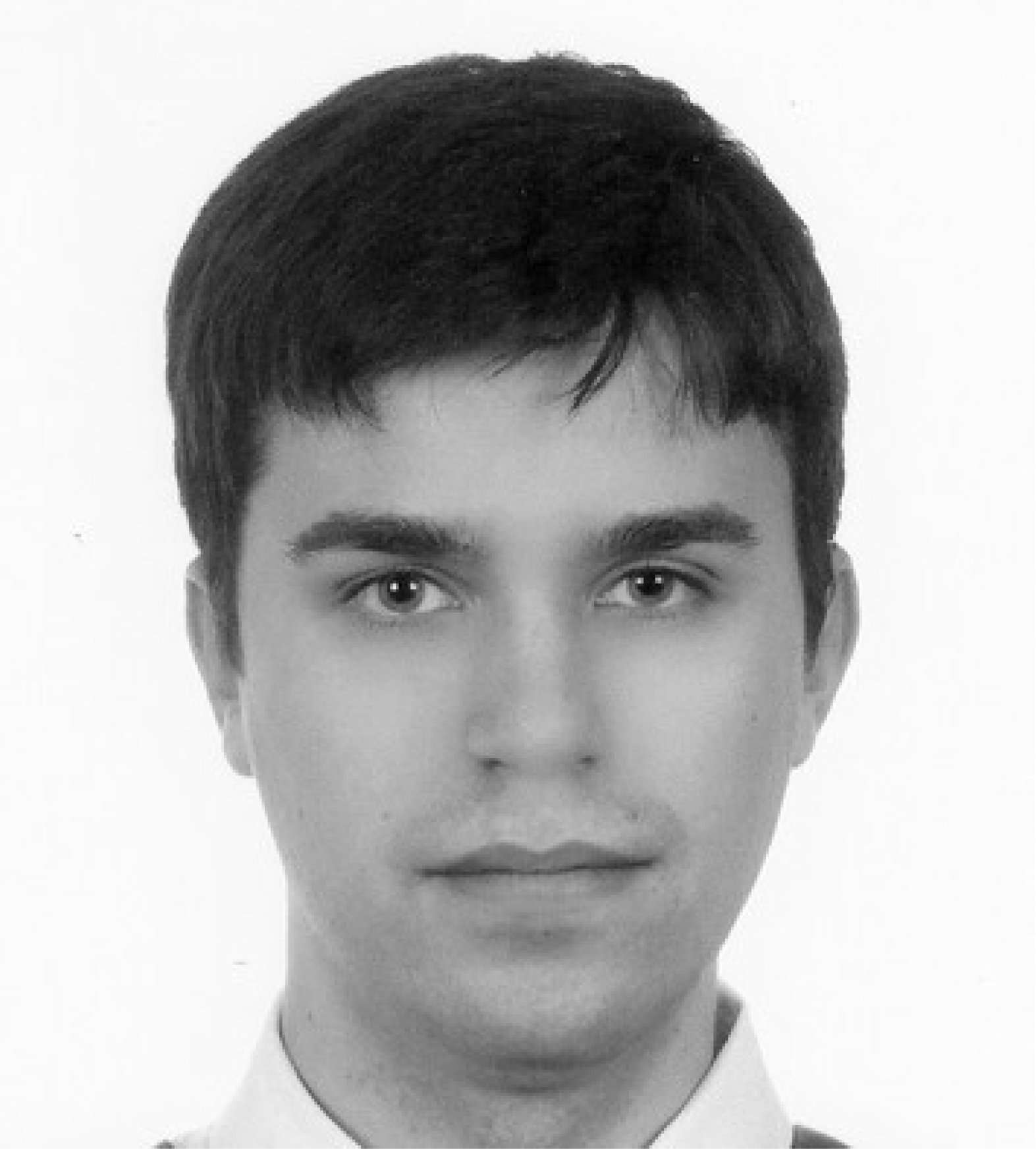}}]{Iv\'an P\'erez}
  is a PhD student in Computer Architecture in the University of Cantabria,
  Spain. He received the B. Sc. degree in Telecomunication Engineer form the
  University of Cantabria in 2014. His research interests include networks on
  chip and memory hierarchy of manycore processors.
\end{IEEEbiography}

\vskip -2\baselineskip plus -1fil

\begin{IEEEbiography}[{\includegraphics[width=1in,height=1.25in,clip,keepaspectratio]{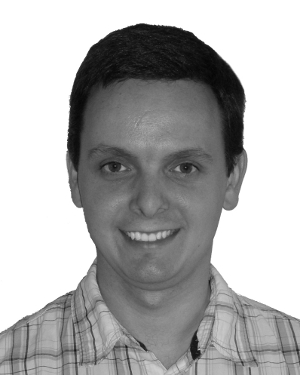}}]{Enrique Vallejo}
	received the PhD degree in computer architecture from the University of Cantabria, in 2010. He is an assistant professor with the University of Cantabria, where he lectures Interconnection Networks in the Computer Science studies. His research interests cover different areas of parallel computing: Interconnection networks, processor microarchitecture, transactional memory, and synchronization mechanisms.
\end{IEEEbiography}

\vskip -2\baselineskip plus -1fil

\begin{IEEEbiography}[{\includegraphics[width=1in,height=1.25in,clip,keepaspectratio]{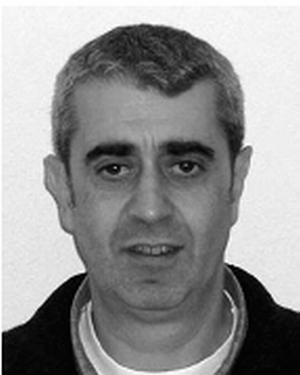}}]{Ram\'on Beivide}
received a PhD degree in computer science and engineering from the Universidad
  Politecnica de Catalunya (UPC), Barcelona, in 1985. He has been an associate
  professor at UPC and the Universidad del Pais Vasco. In 1991, he joined the
  Universidad de Cantabria in Santander, Spain, where he is a full professor of
  telecommunication and computer engineering and he has served as the dean of
  the School of Computer Science. His research interests include parallel
  computers,
  interconnection networks, memory hierarchies and graph theory. He has
  published more than 100 technical papers on such topics.
\end{IEEEbiography}

\end{document}